\documentclass[a4paper]{article}
\pdfoutput=1
\usepackage{pdfpages}

\usepackage{icrc2013}
\usepackage{amsmath}
\usepackage[english]{babel}
\usepackage{textcomp}
\usepackage{url}
\usepackage{hyperref}
\hypersetup{
  colorlinks=true,
  linkcolor=blue,
  citecolor=blue,
  urlcolor=blue,
}

\newcommand{\nosection}[1]{%
  \refstepcounter{section}%
  \addcontentsline{toc}{section}{\protect\numberline{\thesection}#1}%
  \markright{#1}}

\title{The Design, Calibration, and Operation of HAWC: Contributions to ICRC 2013}

\authors{
{\bf The HAWC Collaboration:}\\
A.~U.~Abeysekara$^{a}$,
R.~Alfaro$^{b}$,
C.~Alvarez$^{c}$,
J.~D.~{\'A}lvarez$^{d}$,
R.~Arceo$^{c}$,
J.~C.~Arteaga-Vel{\'a}zquez$^{d}$,
H.~A.~Ayala Solares$^{e}$,
A.~S.~Barber$^{f}$,
B.~M.~Baughman$^{g}$,
N.~Bautista-Elivar$^{h}$
% ...
}

\afiliations{
{\center
$^{a}$ Department of Physics and Astronomy, Michigan State University, East Lansing, MI, USA\\
$^{b}$ Instituto de F{\'\i}sica, Universidad Nacional Aut{\'o}noma de M{\'e}xico, Mexico D.F., Mexico\\
$^{c}$ CEFyMAP, Universidad Aut{\'o}noma de Chiapas, Tuxtla Guti{\'e}rrez, Chiapas, Mexico\\
$^{d}$ Universidad Michoacana de San Nicol{\'a}s de Hidalgo, Morelia, Mexico\\
$^{e}$ Department of Physics, Michigan Technological University, Houghton, MI, USA\\
$^{f}$ Department of Physics and Astronomy, University of Utah, Salt Lake City, UT, USA\\
$^{g}$ Department of Physics, University of Maryland, College Park, MD, USA\\
$^{h}$ Universidad Politecnica de Pachuca, Pachuca, Hgo, Mexico\\
% ...
}
}

\abstract{
  The High-Altitude Water Cherenkov Gamma Ray Observatory (HAWC) is under
  construction 4100 meters above sea level at Sierra Negra, Mexico.  We
  describe the design and cabling of the detector, the characterization of the
  photomultipliers, and the timing calibration system.  We also outline
  a next-generation detector based on the water Cherenkov technique.
}


\begin{document}
\maketitle

\pagestyle{plain}
\pagenumbering{arabic}

\clearpage

% Contents
\newpage
\onecolumn{
  \tableofcontents
}

% Cabling
\newpage
\setcounter{section}{0}
\nosection{HV Cable Manufacture and Testing for the HAWC observatory\\
{\footnotesize\sc Shannon Adams, Ahron~S. Barber, Andrew Fullmer, David Kieda,
Michael Newbold, Ian Sohl, R.~Wayne Springer}}
\setcounter{section}{0}
\setcounter{figure}{0}
\setcounter{table}{0}
\setcounter{equation}{0}
%%
% 33nd International Cosmic Ray Conference - 2013 - Rio de Janeiro, Brazil
% Template adapted from the 2011 ICRC template.
%
%\documentclass[a4paper]{article}
%
%\usepackage{icrc2013}
%\usepackage[english]{babel}

%The paper title
\title{HV Cable Manufacture and Testing for the HAWC observatory}

%The short title to appear at the header of the pages.
\shorttitle{HV Cable Manufacture and Testing}

%All paper authors
\authors{
S. Adams$^{1}$, 
A. S. Barber*$^{1}$,
A. Fullmer$^{1}$,
D. Kieda$^{1}$,
M. D. Newbold$^{1}$,
I. Sohl$^{1}$,
R. W. Springer$^{1}$,

for the HAWC Collaboration$^{2}$.
}
%Shannon Adams, Ahron S. Barber, Alexander Fullmer, David Kieda, Michael Newbold, Ian Shol, [Robert] Wayne Springer 

%All the affiliations.
\afiliations{
$^1$ Department of Physics and Astronomy, University of Utah \\
$^{2}$ For a complete author list, see the special section of these proceedings \\

\scriptsize{
}
}

%email address of the contact person
\email{ahron.barber@utah.edu}

%The abstract.
\abstract{HAWC, the High Altitude Water Cherenkov gamma- and cosmic-ray observatory is being constructed on Sierra Negra in the Pico de Orizaba National Park in Mexico. HAWC is an array of 300 water Cherenkov detectors spread over 22,000 m$^2$. Each detector has four photo multiplier tubes (PMTs). Individual high voltage power to each PMT is supplied from the central electronics facility, and travels to each PMT through 149-meter long RG-59 cables. Signals from each PMT travel back to the facility through the same high voltage cable, and the PMT signal is picked off the HV signal by the front-end board electronics for amplification, triggering, and event reconstruction. Reliable operation of HAWC requires testing of individual cables for the ability to handle the required PMT high voltages. Accurate reconstruction of extensive air shower direction and energy require the equalization of cable time delays as well as the accurate measurement of cable attenuation characteristics, including cable rise time and fall time. In this poster, we describe the fabrication, construction, and deployment of the HAWC cabling system. We outline the cable fabrication procedure used to equalize cable delays to ensure similar attenuation between cables, and describe the automated testing procedures used to certify individual cable high voltage and bandwidth performance. We summarize the characteristics of the cables produced for HAWC, including the average and standard deviation of the cable delays, rise and fall times, and cable attenuation for fast pulses.}

%The keywords

%\begin{document}
\maketitle

%Introduction
\section*{Introduction}

The High Altitude Water Cherenkov Observatory \cite{bib:Miguel} is currently being built in central Mexico near Puebla within the Pico de Orizaba National Park. This is the second generation of water Cherenkov style detectors for gamma and cosmic rays. HAWC is composed of 300 water Cherenkov detectors built over an area of 22,000 m$^2$. Each of these detectors will be instrumented with one 25.5 cm photo multiplier tube (PMT) and three 20.3 cm PMTs. A total  of 1200 PMTs will be  instrumented and cabled to the central electronics facility.  The electronics are centrally located  to minimize the cable length  to the most distant PMTs. Each PMT will have a single  cable; the high voltage and the PMT signals propagate along the same coaxial cable. The PMT signal is isolated by the high voltage section of the front-end-board electronics, and is recorded using a multi-level discriminator and multi-hit TDC.  The dual use of the cable for both high voltage and signal measurement requires properly prepared cables. 

HAWC uses the RG-59 Belden 8241 coaxial cable, which is purchased in spools of $\sim$150 m in length. The core is solid-core copper coated steel with polyethylene insulator and copper braided shield with an outer PVC covering. This cable is designed to operate at a maximum voltage of 1700V RMS with no more than 470 watts at 50 MHz. However, our cables are subjected to 3100V DC in testing. This cable type was successfully used for the outriggers in the MILAGRO experiment, HAWCÕs predecessor. Each cable costs $\sim150$ USD.   

It is important for the cables to be made as electrically identical to each other as possible because HAWC requires accurate timing and similar attenuation for all the cables. On-site timing calibrations will be performed periodically with a laser calibration system to account for the electronics pathways in addition to the cables \cite{bib:Hao}. The accurate reconstruction of the extensive air showers requires precise timing and charge measurements from the PMT signals.  Small changes in the manufacturing conditions at the cable manufacturing factory can affect the electrical properties of the cable. To fabricate cables  with consistent electrical properties, each cable is time matched  to a standard reference  (the ``Golden Cable''). The standard ÔGolden CableÕ was fabricated  for constructing the first batch of cables for the first deployment phase of the HAWC detector: the first 30 detectors \cite{bib:Ibrahim}.

The raw cables are purchased in batches of $\sim$300 spools of 150m length each,  and then processed at the University of Utah before being shipped to the HAWC site in Mexico for installation. In section 2, the fabrication of the cables is discussed. Section 3 discusses the methods used to test and verify the electrical performance of the cables. Distributions of various electrical parameters for the cables, such as time delay, rise and fall times are  presented in section 4.

%Manufacture
\section*{Processing}

The cable fabrication is performed in four steps:
\begin{enumerate}
\item Re-spool the raw cable. 
\item Crimp on first SHV connector
\item Cut the cable length to time match the new cable to the Golden Cable
\item Crimp the second connector
\end{enumerate}

Re-spooling the cables is done for two reasons: to identify any physical damage such as damage to the outer PVC jacket, and to expose a  9 m segment of cable to affix an SHV\footnote{Secure High Voltage} connector and allow electrical testing.  The rate of defective cables, physical damage, is found to be $\sim$1\%. Most defects appear as an abrasion to the PVC outer jacket, or  the outer PVC jacket is split lengthwise. It is important to identify defects in the outer PVC jacket as the cables are routed between the detectors in drainage pipes buried deep enough to avoid night to day temperature variations. These pipes occasionally fill with water, and so these cables must be able to carry the required high voltage even when submerged. 

The cables must have the similar lengths to provide identical time delays, but the distance from the central electronics facility to each tank is different. For shorter distances,  the excess PMT cable will be stored near its specific water tank, wound  on their original spool. The exposed short 9m cable segment connects to the inner windings of the cable on the spool, and therefore allows the  electrical connection to be made to this end of the cable without having to fully unwind the cable off the spool. 

Once a cable has been checked for physical defects and rewound onto the spool with the exposed 9m cable segment from the inner windings, a Kings 1705-1 SHV connector is crimped to the end of the exposed 9m cable segment.   The  other end of the cable is then cut to give precisely the same time delay as the Golden Cable, using a reflection-cancellation technique.

\begin{figure*}[!t]
  \centering
  \includegraphics[width=0.8\textwidth]{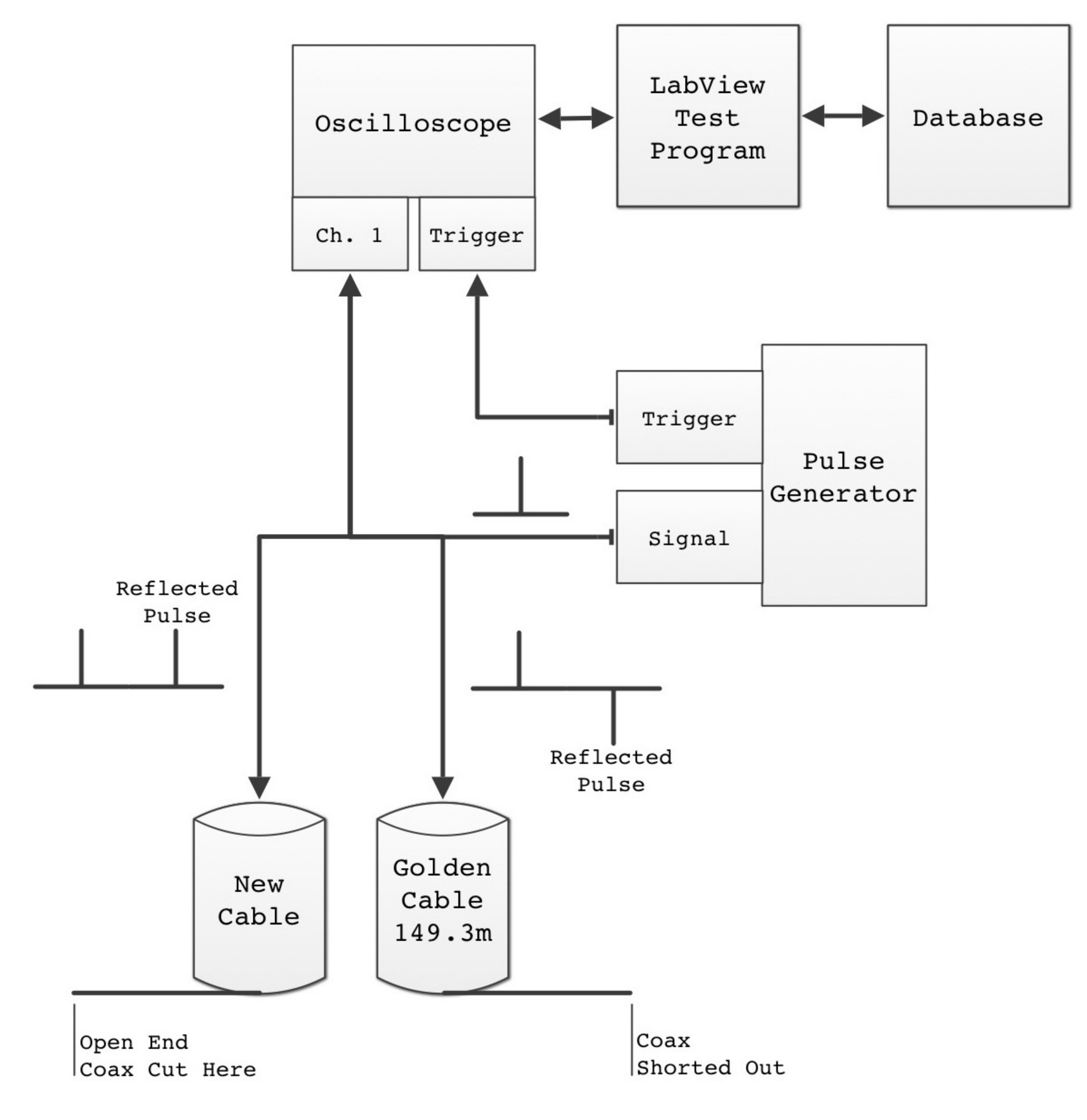}
  \caption{This figure illustrates the connections for the Pulse Generator $(PG)$ which the signal is then split 3 ways, to the Golden Cable $(GC)$ and the Cable $( C )$ and the Oscilloscope $(OS)$. This circuit is a simple 4-way junction containing equidistant paths for the two cables being compared.}
  \label{TimeMatch}
 \end{figure*}

The reflection-cancellation technique employs a fast pulse generator to send a signal through both the Golden Cable and the cable under fabrication (figure \ref{TimeMatch}).  A coaxial 4-way junction is used to split the pulse generator signal and pass it to one end of the Golden Cable, one end of the cable being fabricated, and channel 1 of a Tektronix TDS 580D Oscilloscope. The other end of the Golden Cable has the coaxial cable terminated to a zero Ohm short circuit. The pulse will therefore reflect off the short circuit and return to the oscilloscope with inverted polarity. The other end of the cable under fabrication is left open circuit so the reflected pulse will return without inversion. The sum of the two  reflected pulses are observed to exactly cancel when both cables have the same time delay. This process has the advantage that it is very robust with respect  to recalibration or temperature drift of the oscilloscope time base.  After the cables are time matched, a second Kings 1705-1 SHV connector is crimped onto the open end of the cable being fabricated. . The cable is then assigned a cable serial number, and a  PMT and tank label. 

%Testing
\section*{Testing}

After attaching the second SHV connector, the cables are tested  and the electrical properties of the cable are stored in an excel spreadsheet. The first test is a simple continuity test, which ensures the connectors have been attached properly and are making contact with the coaxial cable. The test is performed with a digital ohmmeter, and the range of acceptable values is listed in table \ref{table_res}. 

\begin{table}[h]
\begin{center}
\begin{tabular}{|l|c|}
\hline Measurement & Acceptable values \\ \hline
Core to Core   & $23 \pm 1 \Omega$ \\ \hline
Shield to Shield   & $1.4 \pm 0.2 \Omega$  \\ \hline
Core to Shield & $\infty$  \\ \hline
\end{tabular}
\caption{The three continuity checks with acceptance bounds.}
\label{table_res}
\end{center}
\end{table}

The second test involves the measurement of the cable performance while subjected to high voltage. The cables are tested at a high voltage of 3100 V. This voltage substantially exceeds the normal high voltage used for the HAWC PMTs (by approximately a factor of two). This test employs a  Bertan 375x, $\pm$5 kV  supply with a current limit set to 1 mA. The power supply is configured for positive high voltage. The high voltage supply is connected through a 10 kOhm inline resistor to one end of the fabricated cable; the other end of the fabricated cable is left open circuit. The voltage across the inline resistor is divided by the 10 kOhm resistance in order to determine the current being supplied by the power supply to the coaxial cable. The high voltage is ramped to 3100V and the voltage across the inline resistor is measured by a handheld voltmeter in order to estimate the current being supplied to the coaxial cable. Cables which pull less than 140 microamps at 3100 V bias are considered acceptable for use in HAWC. 
 
The final test stage is the electrical characterization test. This test uses a BNC6040 pulse generator to send a calibrated square pulse into the cable. The other end of the cable is terminated to 75 Ohms at channel 1 of a LeCroy LC534A digitizing oscilloscope. The LeCroy LC534A oscilloscope measures the various electrical properties of the cable, including rise time, fall time, time delay , and pulse area. These values are then read into the LabView program to apply the acceptance cuts shown in table \ref{table_acce}. This program displays all the data for a batch of cables and an indicator whether the cables passed or failed. The resulting values for each cable are stored in an Excel spreadsheet. 

\begin{table}[t]
\begin{center}
\begin{tabular}{|l|c|c|}
\hline Measurement & High bound & Low bound \\ \hline
Rise Time (ns) & 46 & 48.5 \\ \hline
Fall Time (ns) & 46 & 49.8 \\ \hline
Delay Time (ns) & 760 & 762.5 \\ \hline
Area (ns$\cdot$V) & 181 & 191 \\ \hline
\end{tabular}
\caption{The acceptance bounds for the pass or failure for all cables.}
\label{table_acce}
\end{center}
\end{table}

\section*{Results}

\begin{figure}[h]
  \centering
  \includegraphics[width=0.45\textwidth]{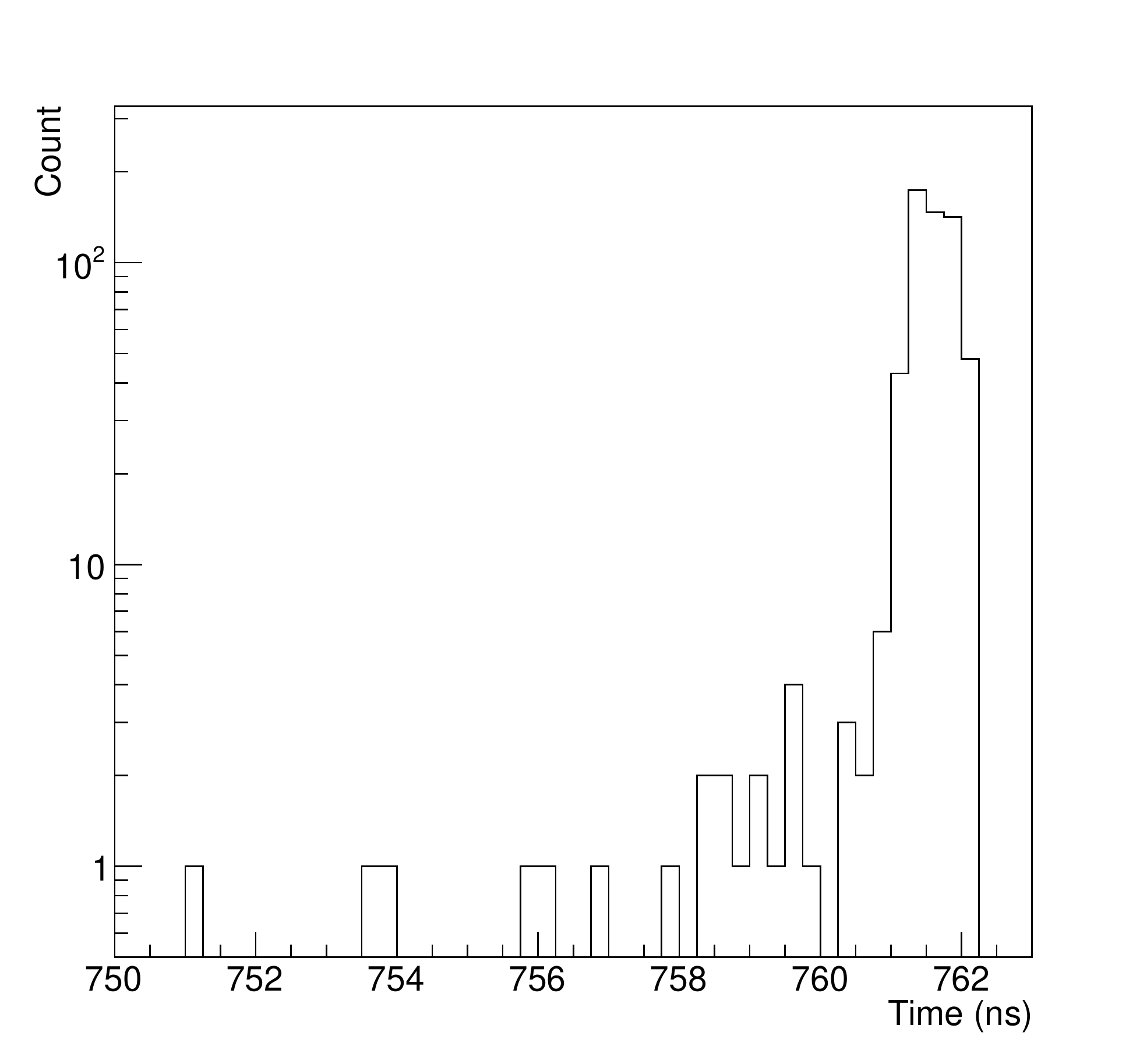}
  \caption{The observed distribution of delay times for 585 Cables. }
  \label{DelayTime}
 \end{figure}

\begin{figure}[h]
  \centering
  \includegraphics[width=0.45\textwidth]{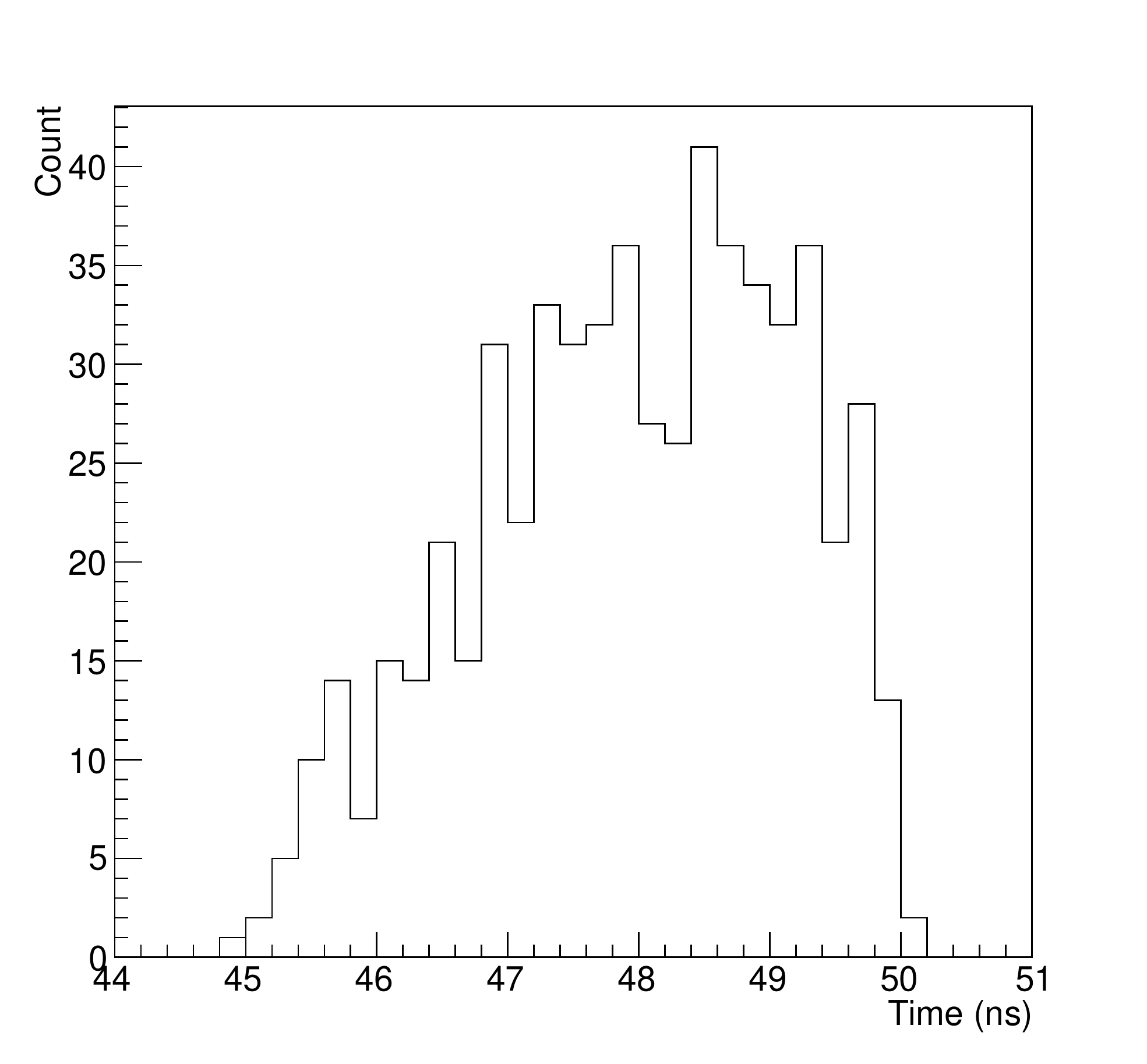}
  \caption{The observed distribution of rise time distribution for 585 cables.}
  \label{RiseTime}
 \end{figure}

\begin{figure}[h]
  \centering
  \includegraphics[width=0.45\textwidth]{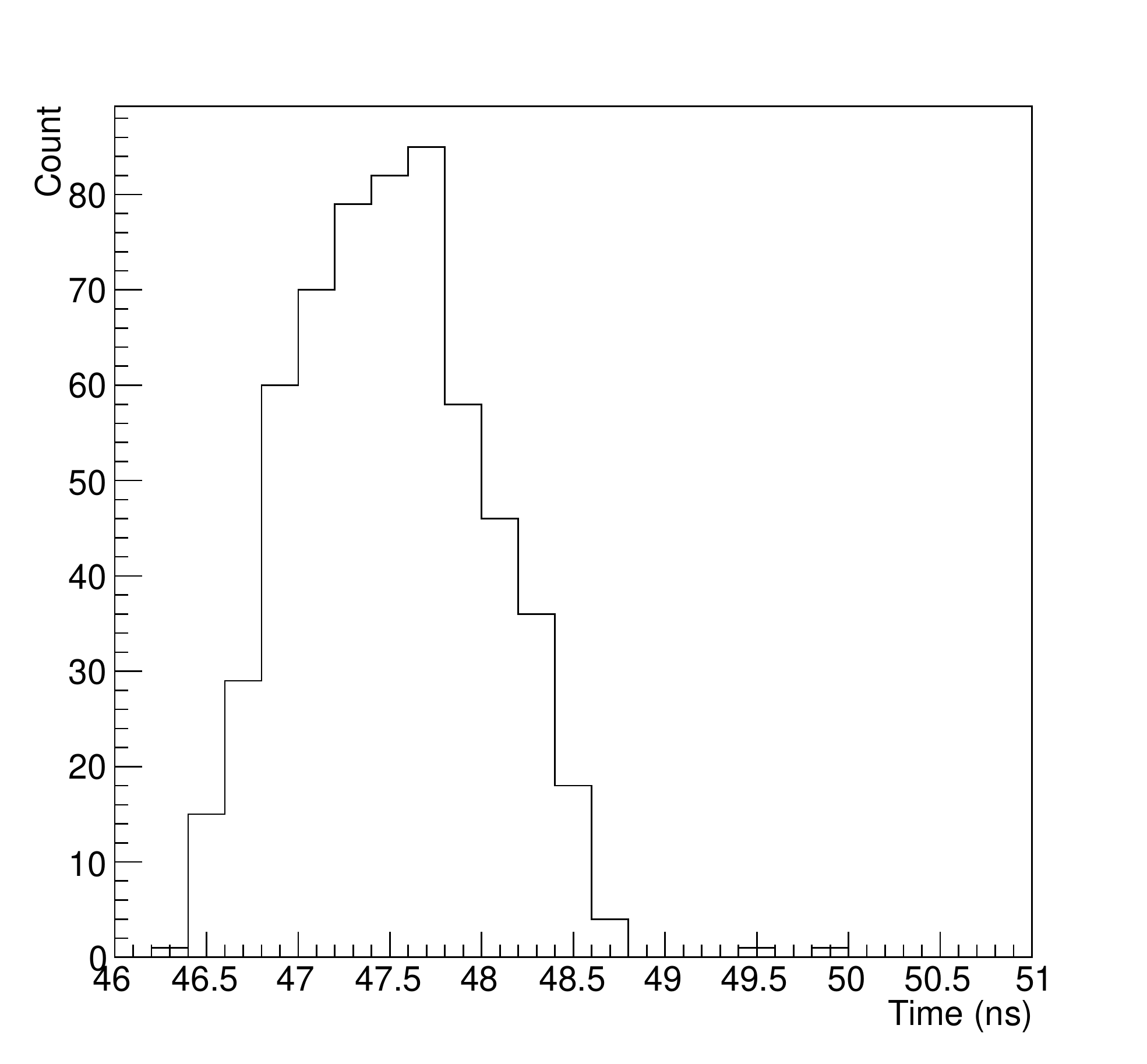}
  \caption{The observed distribution of fall time distribution for the 585 cables.}
  \label{FallTime}
 \end{figure}

\begin{figure}[h]
  \centering
  \includegraphics[width=0.45\textwidth]{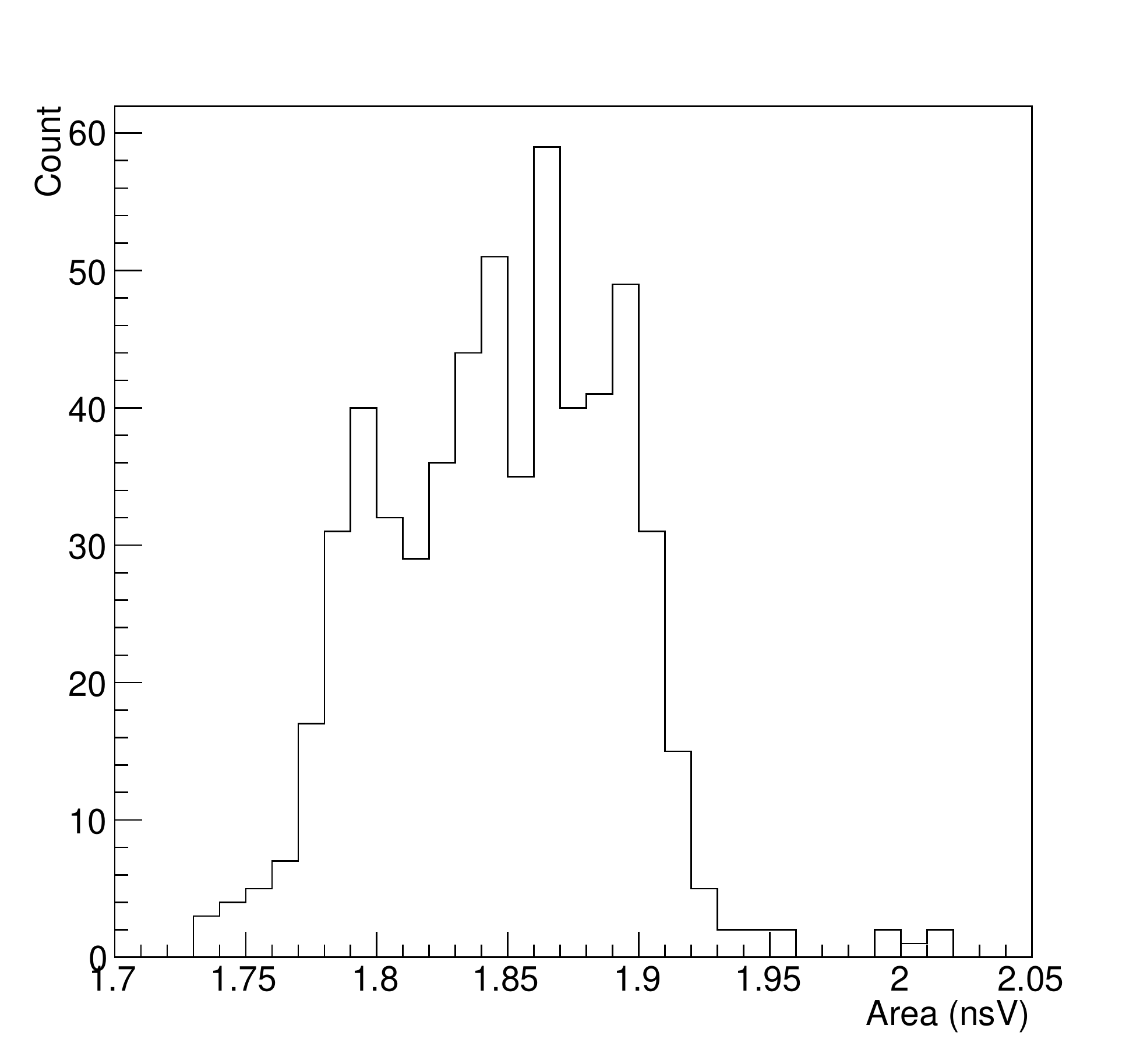}
  \caption{The observed distribution of Area distribution for 585 cables.}
  \label{Area}
 \end{figure}

 The objective of this work is to create a very uniform set of cables for the HAWC experiment. This objective has been achieved, as shown in the table \ref{table_end} along with the figures \ref{DelayTime} through \ref{Area}. Of the final production of cables, we find that there is a failure rate of 3\%. If a cable fails any of the final test value ranges, it is rechecked for manufacturing errors (e.g. a bad SHV crimp). If it still fails, then it is reserved as a spare. Of the failures, 19 cables are found to have too short a time delay and are therefore kept as spares, meaning they do not receive a tank and PMT location until necessary. The obviously short cables are found early in the processing pipeline, as the physical length is found to be shorter than the Golden Cable during the re-spooling. Others are found when the cable is time matched; the reflected pulse is observed to be less than the Golden Cable's nominal time. These not so obvious short cables are then set aside to be spare cables and then processed after all other cables. 
 
In figure \ref{RiseTime}, there are a few cables that are observed below the lower acceptance bound from table \ref{table_acce}. However, as one can see in the figure, the distribution is continuous. Only two of the cables with a delay time less than 760 ns have a rise time faster than 47 ns. The limits set for the cables are arbitrary, and were chosen to highlight those cables falling on the extremes of a predicted distribution. We conclude that for any cable that falls outside the set limits (except for the delay time measurement) and are found to belong within the continuous distribution are acceptable. The fall time and the area both have a few outlying cables, but are within the chosen acceptable range or deemed acceptable. 

\section*{Acknowledgments}

US National Science Foundation (NSF); US Department of Energy Office of
High-Energy Physics; The Laboratory Directed Research and Development (LDRD)
program of Los Alamos National Laboratory; Consejo Nacional de Ciencia y
Tecnolog\'{\i}a (CONACyT), M\'exico; Red de F\'{\i}sica de Altas Energ\'{\i}as,
M\'exico; DGAPA-UNAM, M\'exico; and the University of Wisconsin Alumni Research
Foundation.

%\vspace{1cm}

\begin{table}[t]
\begin{center}
\begin{tabular}{|l|c|c|}
\hline Measurement & Average & Standard Deviation \\ \hline
Rise Time (ns)  & 47.9 & 1.2 \\ \hline
Fall Time (ns) & 47.5 & 0.5 \\ \hline
Delay Time (ns) & 761.2 & 0.8 \\ \hline
Area (ns$\cdot$V) & 184.8 & 4.5 \\ \hline
\end{tabular}
\caption{This is the end result for the 300 cables in the second and third batches of cables.}
\label{table_end}
\end{center}
\end{table}

%Plot of the delay times

%Plot of the Rise times
 
%Plot of the Fall Time

%Plot of the Area

\clearpage

%\end{document}

% R5912 Scanning
\newpage
\setcounter{section}{1}
\nosection{Characterization of the HAWC R5912 Photomultipliers\\
{\footnotesize\sc Pablo Vanegas, Rosalia Langarica, Gerardo Lara, Luis~A.
Mart\'{i}nez, Silvio Tinoco, Ruben Alfaro, Arturo Iriarte, Andr\'{e}s Sandoval}}
\setcounter{section}{0}
\setcounter{figure}{0}
\setcounter{table}{0}
\setcounter{equation}{0}
%%
% 33nd International Cosmic Ray Conference - 2013 - Rio de Janeiro, Brazil
% Template adapted from the 2011 ICRC template.
%
%\documentclass[a4paper]{article}
%
%\usepackage{icrc2013}
%\usepackage[english]{babel}

%The paper title
\title{Characterization of the HAWC R5912 photomultipliers}

%The short title to appear at the header of the pages.
\shorttitle{Third Generation WCD}

%All paper authors
\authors{
P. Vanegas$^{1}$
R. Langarica$^{2}$
G. Lara$^{2}$
L. A. Mart\'inez$^{2}$
S. Tinoco$^{2}$
R. Alfaro, $^{1}$
A. Iriarte$^{2}$ 
A. Sandoval$^{1}$,

for the HAWC Collaboration$^{3}$.
}

%All the affiliations.
\afiliations{
$^1$ Instituto de F\'{\i}sica, Universidad Nacional Autonoma de M\'exico, M\'exico City \\
$^2$ Instituto de Astronom\'{\i}a, Universidad Nacional Autonoma de M\'exico, M\'exico City 
$^3$ For a complete author list, see the special section of these proceedings
}

%email address of the contact person
\email{asandoval@fisica.unam.mx}

%The abstract.
\abstract{Each of the 300 water Cherenkov detectors of the HAWC Observatory is instrumented with three 8-inch Hamamatsu R5912 photomultiplier tubes (PMTs) and a central 10-inch Hamamatsu R7081 high quantum efficiency PMT. The 900 R5912 PMTs are reused from the Milagro experiment, which was operated from 2000 to 2008 in the Jemez Mountains near Los Alamos, NM. In order to characterise the present performance of these PMTs, an automatic system was designed and constructed to scan the photocathode surface with a pulsed light source to determine the gain variations at 101 points distributed over the active surface. Gain and efficiency variations over the photocathode surface can be studied with large pulses and with single photo electrons. We find systematic non-uniformities that seem related to the geometry of the entry window into the dynodes.}

%The keywords

%\begin{document}
\maketitle

%Begin a section.
\section*{Introduction}
The HAWC Collaboration is building a high-energy gamma ray observatory at a latitude of 19 deg north and 4,100 m asl in the Sierra Negra mountain in the State of Puebla, central Mexico \cite{bib:hawc}. The array will consist of 300 water Cherenkov detectors (WCD) that will study the incoming gamma/cosmic rays in the hundred of GeV to hundreds of TeV energy range. This is done through the detection of the Cherenkov light emitted by the related air shower particles as they enter the large water containers. 

HAWC will reuse the 900 Hamamatsu R5912 photomultipliers of 20 cm diameter from Milagro \cite{bib:milagro}. In order to characterise their present
performance we scan the active area of the photocathode in order to measure the efficiency and gain. A
robotic characterisation system (RCS) was designed and manufactured to achieve automated measurements over 101 points
distributed on the PMT active spherical surface. 

\section*{The Robotic Characterisation System}

The RCS is composed of three main subsystems: the mechanical, the electronic and control and the data acquisition subsystems.
There is a mechanisms for positioning a
LED normal to the photocathode surface in a light-tight cabinet. The positioning mechanisms are driven by stepper motors, which along with micro-switches and
cabling compose the electronic subsystem. In addition there is an open-loop control system that includes the software for the
controllers, the definition of home position, the pulsed triggering, etc.

\begin{figure}[t]
  \centering
  \includegraphics[width=0.4\textwidth]{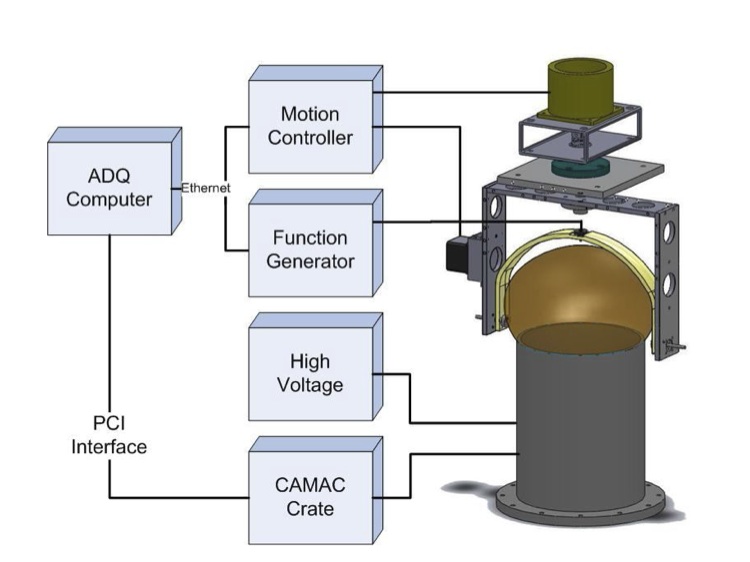}
  \caption{Diagram of the scanning robot with its different functional parts.}
  \label{fig1}
 \end{figure}

The RCS has an arched arm with the inner segment having a 25 cm diameter and a light source at its apex that can be
moved in azimuthal an vertical directions and thereby sweep the entire surface of the photocathode. The light source coupled to a short optical fibre 
produces a narrow beam spot of 3 mm diameter on the photocathode surface.  It is an ultra bright (9000 mcandles) blue LED whose emission is centred at
470 nm, and extends from 440 to 500 nm. The LED is driven by a Tektronix arbitrary function generator (TFG) with a
square pulse of 10 nsec width; the amplitude can be adjusted to get from one single photoelectron (PE) to 50 PEs. The data acquisition
(DAQ) has been done by a CAMAC system (Wiener CC32 PCI controller and a 2249A LeCroy ADC) controlled by a
computer with MS Windows operating system, the same computer controls the TFG. The trigger for the DAQ is taken
from the TFG; this allows measuring the time resolution of the PMT as function of the number of PE and monitoring the
efficiency of the photocathode. Figure \ref{fig1} shows a diagram of the experimental setup.

In order to keep the light source perpendicular to the surface as it scans over the sensitive area the RCS was designed with  two orthogonal rotational axis. 
The vertical axis carries a rigid frame with the second axis that rotates an aluminium arch that is concentric with the PMT surface and which carries the light source at the apex.

\begin{figure}[t]
  \centering
  \includegraphics[width=0.4\textwidth]{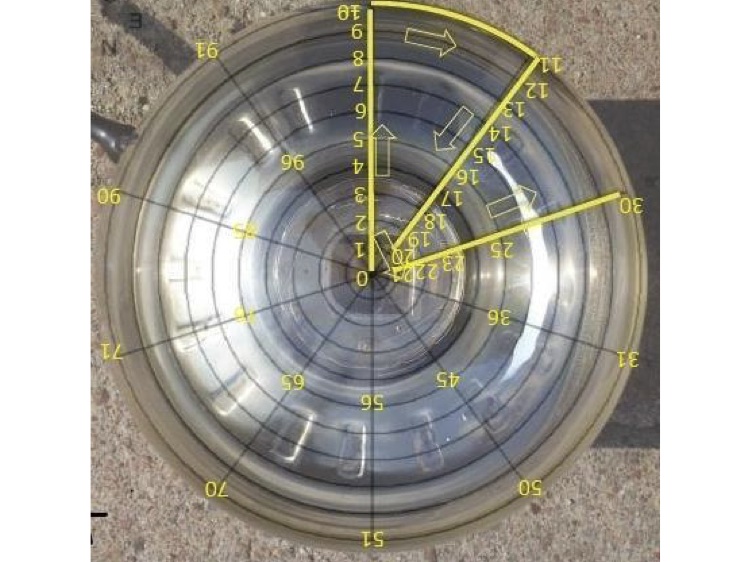}
  \caption{Image of the R5912 PMT with the scanning pattern of the 101 measured points indicated. }
  \label{fig2}
 \end{figure}

A picture of the PMT with the scanning pattern is shown in Fig. 2  \ref{fig2}. One can see the shape of the entrance window to the dynodes in the centre.

The RCS software consists of three modules. The first module interacts with the stepper motor controller by means of a
network connection through a TCP socket. This module allows the user to control azimuthal and elevation stepper
motors and setup their basic operation parameters. The second module is used to control the Tektronix AFG 3252
function generator which in turn sends the signal to the LED. The module uses TeKVISA connectivity library provided
by Tektronix. TekVISA provides communication with instruments via an Ethernet link. The third module is the main
program, a GUI that is in charge of data acquisition. It interacts with the CAMAC library provided by the crate
manufacturer to get data from the LeCroy ADC. Once the user defines the points over the PMT surface where the LED
will be positioned the GUI starts the data acquisition run. At present, the 101 point run takes less than 15 minutes. 

\section*{Results}

\begin{figure}[t]
  \centering
  \includegraphics[width=0.5\textwidth]{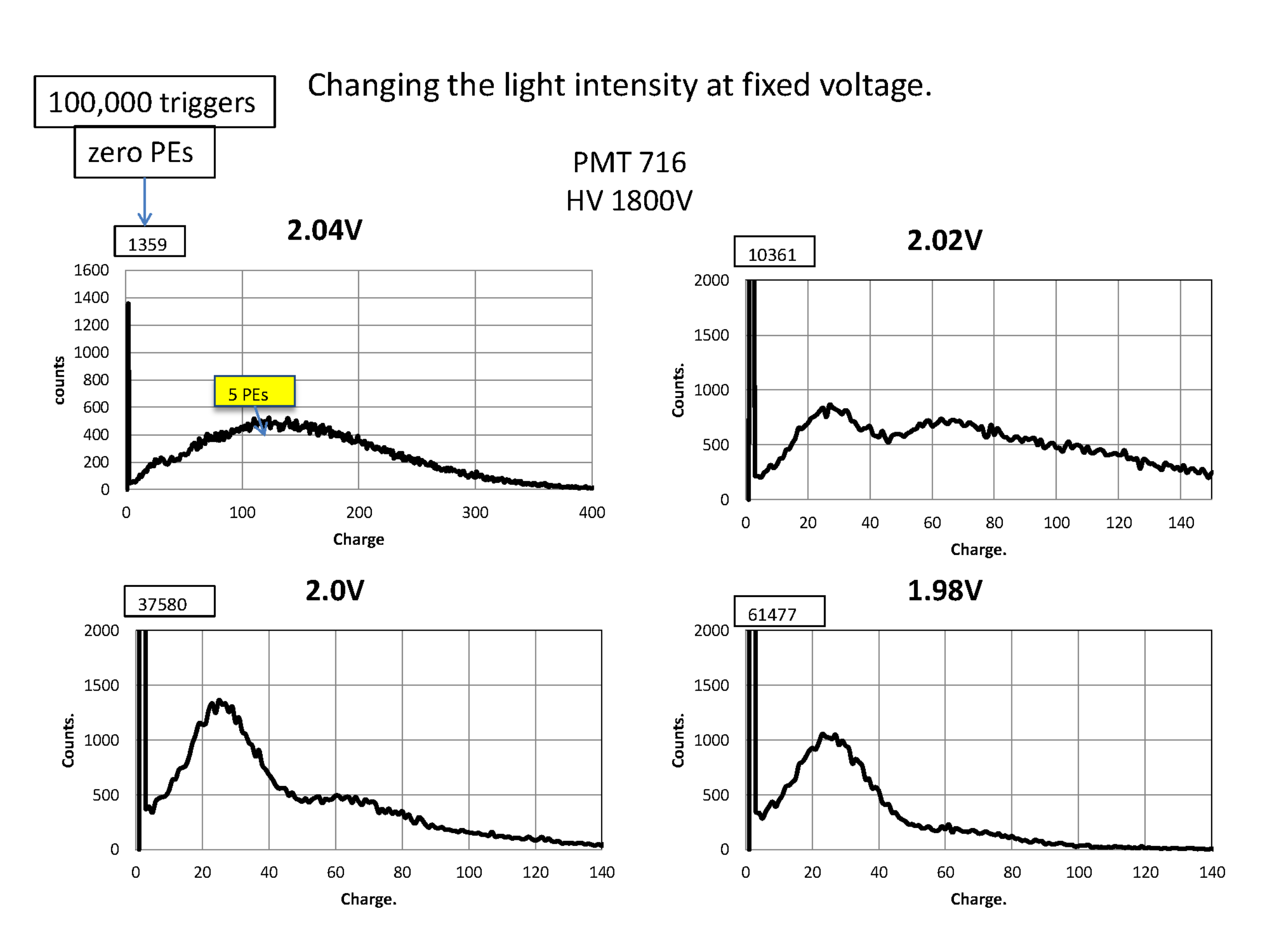}
  \caption{Response of the PMT to different light intensities of the LED.}
  \label{fig3}
 \end{figure}

The pulses from the PMTs as a response to the flashing of the LED are digitised by a charge integrating ADC in 
a 20 ns time window with a conversion of 0.25 pC/channel. For each measurement  the LED is flashed 10,000 times at
a given excitation voltage VLED. The gain curves are measured by varying the HV on the PMT from 1400 V to 1850 V.
Typical operating voltages correspond to a gain of 1.5x$10^7$.  The LED light intensity can be varied by changing the excitation 
voltage provided to the LED by the TFG. Fig \ref{fig3} shows the charge spectra for the LED mean light output varying from 1 to 5 PE.

\begin{figure}[t]
  \centering
  \includegraphics[width=0.5\textwidth]{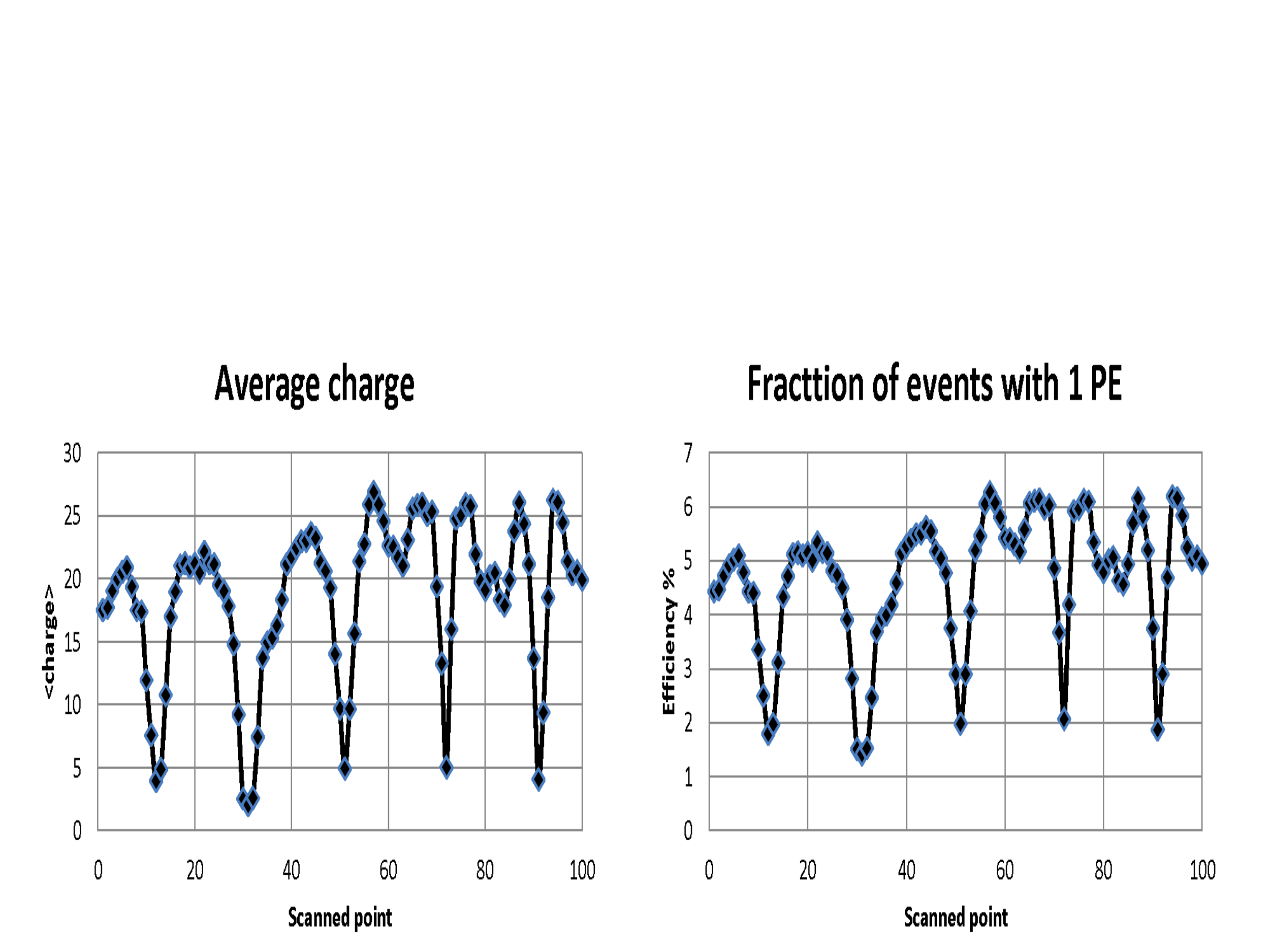}
  \caption{Measured profiles  of the gain variations as the pulsed light source scans the PMT surface.}
  \label{fig4}
 \end{figure}

\begin{figure}[ht]
  \centering
  \includegraphics[width=0.5\textwidth]{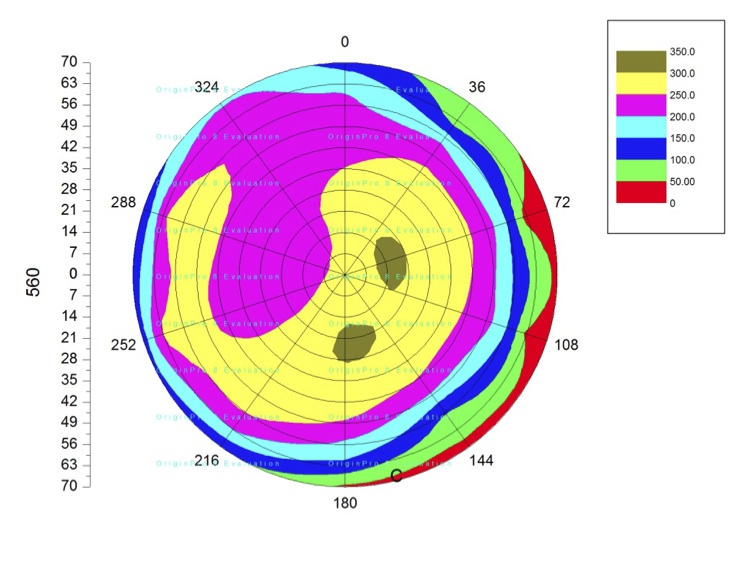}
  \caption{Contour plots of constant gain over the PMT surface.}
  \label{fig5}
 \end{figure}

Profiles of the PMT response to a fixed illumination from the LED flash are shown in Fig. \ref{fig4} for the scanning of the 101 points.
One sees there are inhomogeneities as the light source scans the surface of a given PMT. These are more clearly shown in the contour
plot of Fig \ref{fig5}, which shows the variations of the average charge of the PMT signal as function of the position of the
light source on the photocathode surface. There is a higher gain on the left side of the figure.

\section*{Conclusions}

The Hamamatsu R5912 photomultipliers used in the Milagro experiment will be
reused by the HAWC gamma ray observatory. In order to evaluate the status of
these PMTs after 8 years of use, a robotic characterisation system was
constructed that can scan the photocathode of the PMTs with a pulsed light
source. The average charge of the digitised signal is used to study gain
variations over the photocathode surface. A systematic effect of a lower gain
over the round edge of the window to the dynodes was observed.

\section*{Acknowledgments}

We acknowledge the support from: US National Science Foundation (NSF); US
Department of Energy Office of High-Energy Physics; The Laboratory Directed
Research and Development (LDRD) program of Los Alamos National Laboratory;
Consejo Nacional de Ciencia y Tecnolog\'{\i}a (CONACyT), M\'exico; Red de
F\'{\i}sica de Altas Energ\'{\i}as, M\'exico; DGAPA-UNAM, M\'exico; and the
University of Wisconsin Alumni Research Foundation.

\clearpage

%\end{document}

% ++HAWC
\newpage
\setcounter{section}{2}
\nosection{A Third Generation Water Cherenkov Observatory\\
{\footnotesize\sc Andr\'{e}s Sandoval}}
\setcounter{section}{0}
\setcounter{figure}{0}
\setcounter{table}{0}
\setcounter{equation}{0}
%%
% 33nd International Cosmic Ray Conference - 2013 - Rio de Janeiro, Brazil
% Template adapted from the 2011 ICRC template.
%
%\documentclass[a4paper]{article}
%
%\usepackage{icrc2013}
%\usepackage[english]{babel}

%The paper title
\title{A Third Generation Water Cherenkov Observatory}

%The short title to appear at the header of the pages.
\shorttitle{Third Generation WCD}

%All paper authors
\authors{
A. Sandoval$^{1}$,

for the HAWC Collaboration$^{2}$.
}

%All the affiliations.
\afiliations{
$^1$ Instituto de Fisica, Universidad Nacional Autonoma de Mexico, Mexico City \\
$^2$ For a complete author list, see the special section of these proceedings \\
}

%email address of the contact person
\email{asandoval@fisica.unam.mx}

%The abstract.
\abstract{The construction of the High Altitude Water Cherenkov (HAWC) gamma ray observatory will be completed in 2014. By September of 2013, HAWC will start continuous operations with the first third of the 300-detector array. As the commissioning of the instrument is approaching planning for a third generation Water Cherenkov Observatory can be done. Several ideas to improve the sensitivity, the gamma/hadron discrimination at lower energies and the energy and angular resolution of a continuously operated, large field of view detector array will be discussed. A path to optimize an instrument for the Southern Hemisphere is presented.}

%The keywords

%\begin{document}
\maketitle

%Begin a section.
\section*{General Requirement}
The HAWC Collaboration is building a high energy gamma ray observatory at a latitude of 19 degrees north and 4,100 m altitude asl on the Sierra Negra mountain in the State of Puebla, central Mexico \cite{bib:hawcSensi}. The array consist of 300 water Cherenkov detectors (WCD) instrumented to detect the incoming gamma/cosmic ray through the Cherenkov light emitted by the air shower particles as they traverse the large water containers. The first 30 detectors went into operation in December 2012. One third of the array was constructed by May of this year and will enter continuous scientific operation in the fall of 2013. The full observatory is expected to enter operations late in 2014.

The aim of this presentation is to generate a discussion of the desirability of establishing a similar observatory in the Southern Hemisphere and present some improvements that could be implemented over the HAWC design. These are presented as a Gedanken experiment with little hard facts. Detailed simulations and optimizations need to be done by groups willing to explore such a possibility. The focus is on gamma rays although such an instrument would also simultaneously observe the much more abundant cosmic rays.

Such a continuously operating  instrument with a wide field of view and continuously operating would survey the Southern skies and in particular the center of the Galaxy and would complement HAWC, the lower energy gamma ray  satellite-based observatories like FERMI, SWIFT  as well as other future observatories such as the planned CTA-South air Cherenkov telescopes.

\section*{HAWC design}

The HAWC observatory consists of individual metallic tanks of 7.3 m diameter and 5 m height with a light- and water-tight plastic membrane, the ÒbladderÓ, containing the highly purified water and 4 hemispherical photomultipliers (PMT) anchored to the bottom and looking up to detect the Cherenkov light of the impinging particles.  The large area PMTs have sensitivity to single photo electrons and sub-nanosecond timing resolution. HAWC will cover a 22,000 m2 site with 12,000 m$^{2}$ of active surface.

\begin{figure*}[!t]
  \centering
  \includegraphics[width=0.8\textwidth]{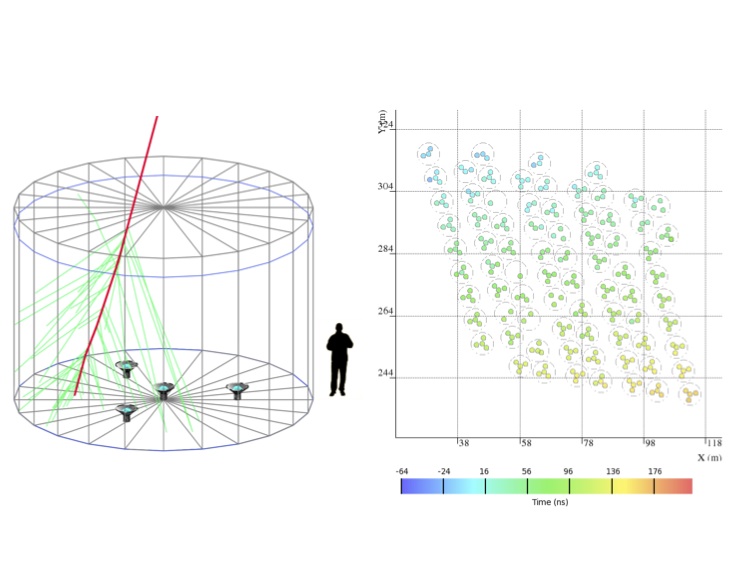}
  \caption{Left a HAWC WCD, right one event of HAWC100, the arrival time is indicated by the colours}
  \label{fig1:wcd}
 \end{figure*}

Consisting of 300 individual water Cherenkov detectors (WCD) it detects the particles from atmospheric showers by their production of Cherenkov light as they enter the 200,000 litre water containers. It can detect showers in the hundred GeV to hundreds of TeVs over a large field of view over the detector and with very high duty cycle. Figure\ref{fig1:wcd} shows a schematic diagram of a HAWC WCD and a measured event from HAWC during the construction phase having 80 WCD.

The direction of the primary particle can be reconstructed from the arrival time of the shower front over the different detectors of the array, their energy is estimated from the number of fired PMTs and their integrated signals. In order to discriminate between primary gamma or cosmic ray, producing an electromagnetic or a hadronic air shower, the topology of the event is used. Electromagnetic showers are smooth with a monotonically decreasing energy density of the shower particles as function of the distance from the shower centre. Hadronic interactions on the other hand produce large angle high momentum particles that produce energy clumps and a larger muon component from the decay in flight of pions and kaons. The performance of the HAWC observatory is discussed in  \cite{bib:pretz}.

\section*{A third generation observatory}

The HAWC detector performance is based on the arrival time and amplitude of the signals at the 4 PMTs of each WCD. Some of the several possible improvements to the 
HAWC design will be discussed below. A detailed cost/performance analysis would need to be done to evaluate what a realistic next generation observatory would look like and estimate itÕs overall capabilities.

\subsection*{Area and Layout}

The sensitivity of a HAWC-like array grows, to first order linearly with the area of the array. An array  2 to 4 times larger than HAWC would array will probably be affordable under current funding scenarios.  For an even  larger array the geometry should be optimized. Namely it should have a more compact central core to improve the detection at the lower energies and get more sparse at larger distances to increase the sensitivity at energy beyond 100 TeV.

\subsection*{Altitude}

By siting the Southern observatory at a higher altitude than the 4,100 m of the HAWC array one would be closer to the shower maximum thereby increasing the lower energy response and improving the sensitivity, the angular and energy resolution and the gamma/hadron discrimination. Places like ALMA in the Atacama Desert of northern Chile have suitable sites at 5,000 m asl. Even higher elevations could be found in Bolivia, but they might be too cold for a WCD.

\subsection*{Detection of the air shower particles}

The detection of the electromagnetic shower particles in a WCD is done trough the Cherenkov light emitted as the particles enter the water tank of the individual detectors. The amount of emitted Cherenkov light can be increased by placing a 1 radiation length layer of lead above the water tanks as a converter. 

\begin{figure}[t]
  \centering
  \includegraphics[width=0.4\textwidth]{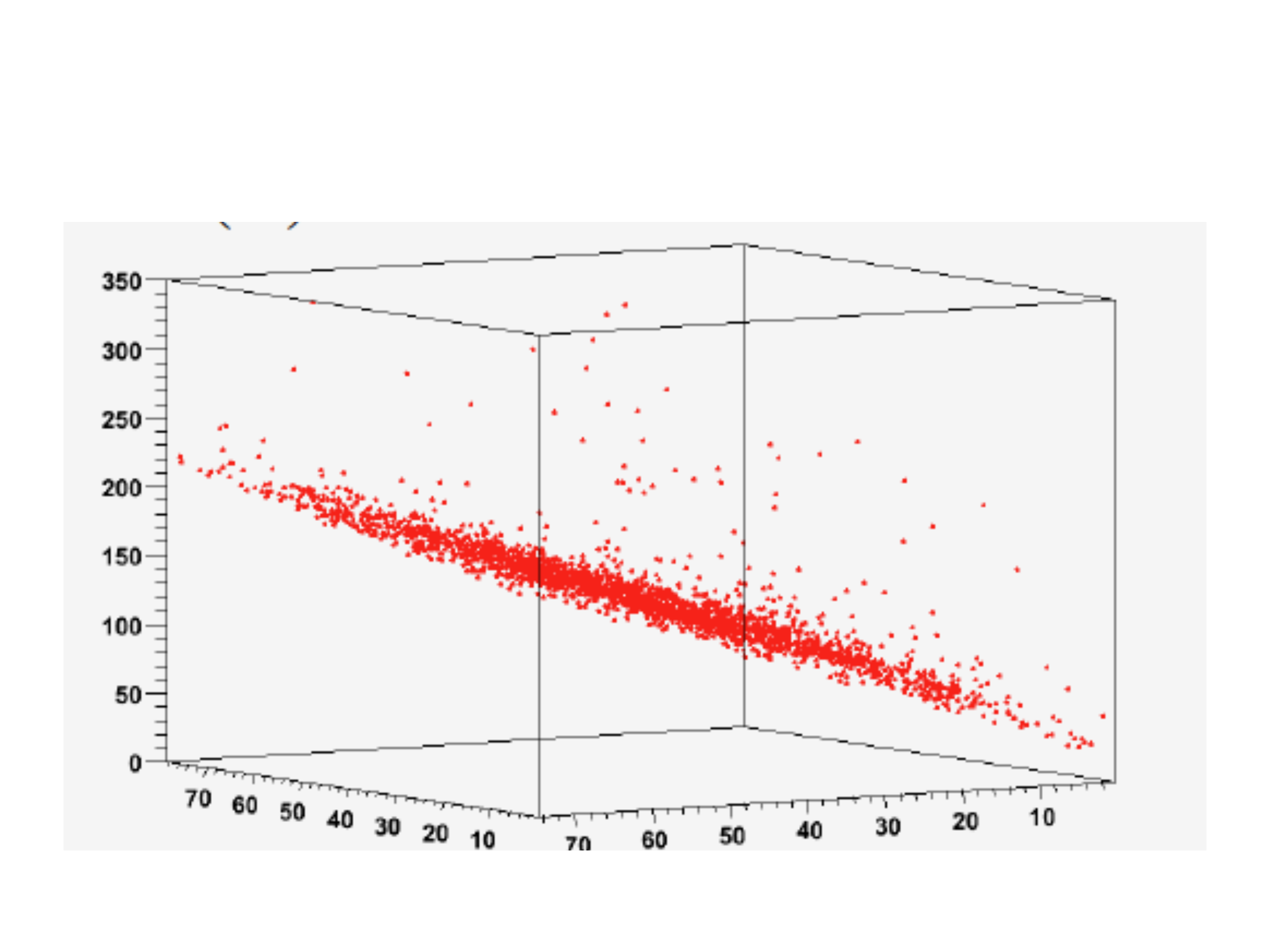}
  \caption{Shower plane measured by the ARGO experiment}
  \label{fig2:showerplane}
 \end{figure}

In a HAWC-like WCD the shape and direction of the incoming shower front is determined by the arrival time of the Cherenkov light to the PMTs in the different WCD detectors. The direction of the incoming primary gamma ray is obtained from the normal to the fitted shower front shape. Although the PMTs have sub-nanosecond time resolution, the fact that each one of them sees a large fraction of the total water volume gives an inherent uncertainty on the position of the point that emits the first detected Cherenkov light that translates in a timing uncertainty.  The ARGO experiment  \cite{bib:ARGO} has shown that it is possible to cover large areas of ground based shower detectors with thin gas detectors having 100 ps time resolution. The segmentation in ARGO into 56 x 46 cm2 pads results in a 1.8 ns time resolution. A finer granularity would be desirable to measure the arrival of the shower front and itÕs width with a 0.5 ns resolution. Figure  \ref{fig2:showerplane} shows the measurement of a shower plane as seen by the ARGO detector.

\begin{figure}[t]
  \centering
  \includegraphics[width=0.4\textwidth]{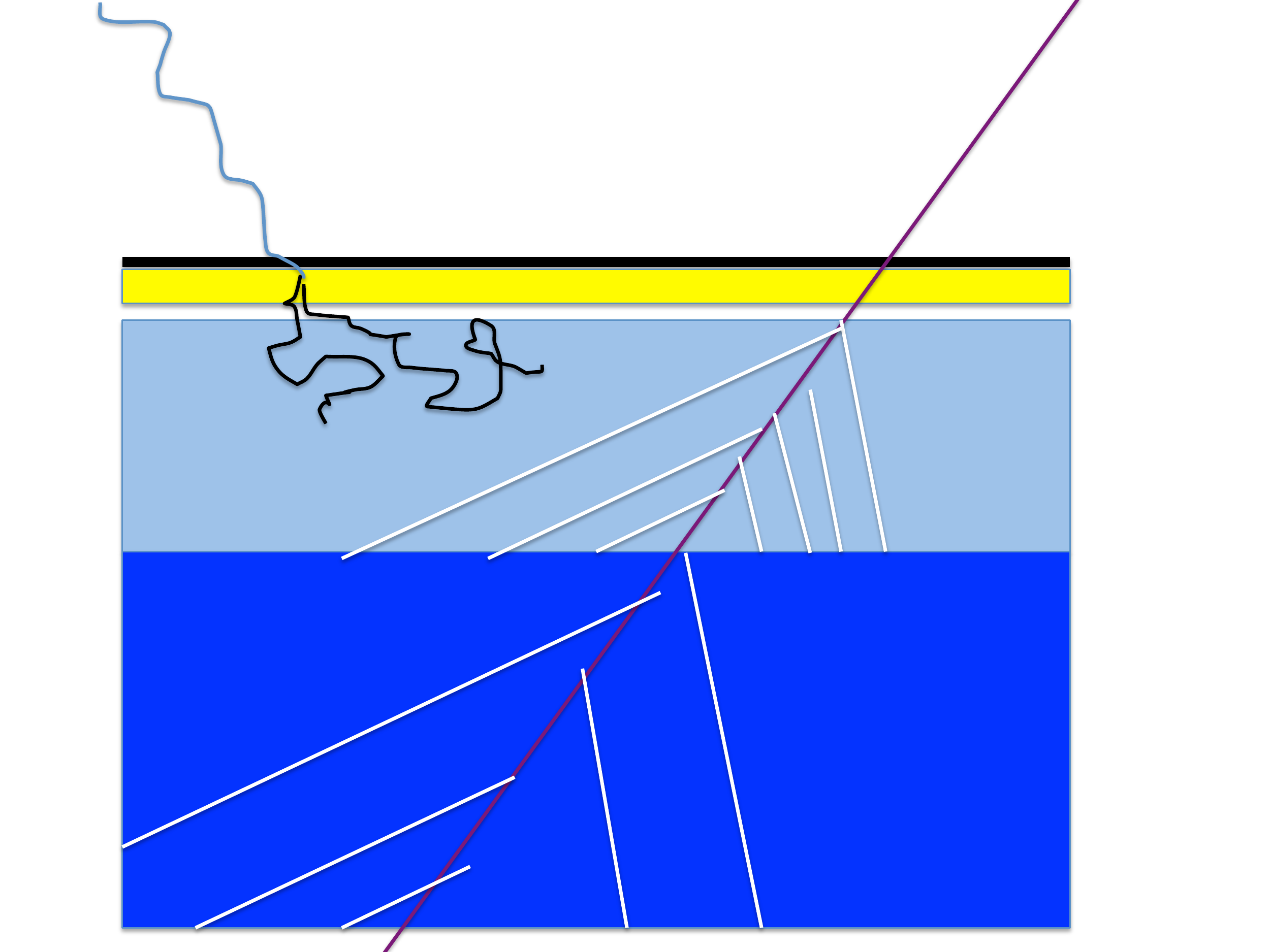}
  \caption{Multi-layered gamma observatory having a timing array and a EM layer and muon layer}
  \label{fig3:layers}
 \end{figure}

The energy of the primary gamma ray is inferred from the signal produced by the PMTs from the detected Cherenkov light at the shower front. The EM shower 
deposits itÕs energy in the top 1.5m water layer. In HAWC the PMTs are placed at the bottom of the tank, 4.5 m from the surface in order to detect also the much larger signals from a through going muon and to improve the gamma/hadron discrimination at the cost of reducing the light seen from EM showers. In a new detector that divides the water volume into two regions  these functions can be separated  optimizing their measurement. A 2 m upper ÒbladderÓ with white reflecting walls, seen by several PMTs would integrate a much larger fraction of the total Cherenkov light from the EM component. A lower 3m deep ÒbladderÓ would detect the through going muon component. In order to shield this volume better, the spaces between tanks could be filled with dirt, effectively burying the tanks. Figure \ref{fig3:layers} shows such a segmented WCD.

The separation of the measurement of the EM component on the upper 2m of the water volume and the muon and hadron signal starting 2m below the surface will improve the gamma/hadron separation by having a better measurement of possible muons into the shower core.

\subsection*{Determination of the shower maximum and improvement of the pointing accuracy}

The inherent fluctuations in the shower development given by the variations in atmospheric depth at which the primary gamma ray interacts thereby initiating the EM cascade effectively degrades the energy resolution in the case where the shower depth is not measured as is the case n HAWC. The current HAWC detector provides no information that could determine the depth of the shower maximum  for each event, but in a new instrument this could be done. Already the shape and with of the shower front determined by the timing array contains information to restrict the range of the  estimate of the depth of the shower maximum. More detailed information could be obtained, at least during night time, by deploying an array of non imaging Cherenkov counters sampling the atmospheric Cherenkov light produced during the development of each shower. Arrays such as AROBIC and HiSCORE  \cite{bib:HISCORE} have demonstrated their performance. In this case to complement the timing and WCD arrays, the non focusing air Cherenkov array should be much more compact than HISCORE. The combination of the timing and air Cherenkov observations have the capability substantially improving the angular resolution and therefore the imaging of the gamma ray sources.

\subsection*{Electronics, trigger and data acquisition}

In the HAWC WCDs, the centrally located high quantum efficiency PMTs have the highest single rates of approximately 50 kHz for signals above one quarter of a single photoelectron. The PMT signals are digitized with a dual threshold time over threshold (ToT) by multi-hit TDCs. The triggers are then performed by a central computer farm in software and stored on disk.

A next generation observatory, with possibly several layers of detectors: a timing array, an array detecting the EM component of the showers, a muon-hadron array and non imaging air Cherenkov detectors would have count rates that can easily be pre-processed and digitized with distributed electronics near the individual detectors. Local groups of detectors can provide level zero triggers which could be processed at higher DAQ level producing more global triggers.

\section*{Conclusions}

It is argued that a it would be desirable to have a wide field of view, continuously operating gamma ray observatory to survey the skies of the Southern Hemisphere complementing the HAWC water Cherenkov array being built at 19 degrees northern latitude at 4,100 m asl in the mountains of central Mexico.

There are several directions to improve over the HAWC layout: increasing the sensitive area, placing it at a higher altitude and separating the detection functions into different layers of the array. These could result in a finely segmented timing array to define the shower front, a layer of water Cherenkov detectors to measure the EM component of the air showers followed by a deeper layer of water Cherenkov detectors measuring the muon and hadron component. Information about the development of the air shower can be obtained during clear nights with an array of non-focusing Cherenkov counters. Information on the depth of the shower maximum improves the energy resolution and the measurement of the arrival times of the air Cherenkov light improves the angular reconstruction of the direction of the primary gamma ray.

\section*{Acknowledgments}

We acknowledge the support from: US National Science Foundation (NSF); US
Department of Energy Office of High-Energy Physics; The Laboratory Directed
Research and Development (LDRD) program of Los Alamos National Laboratory;
Consejo Nacional de Ciencia y Tecnolog\'{\i}a (CONACyT), M\'exico; Red de
F\'{\i}sica de Altas Energ\'{\i}as, M\'exico; DGAPA-UNAM, M\'exico; and the
University of Wisconsin Alumni Research Foundation.

\clearpage

%\end{document}

% HAWC Deployment
\newpage
\setcounter{section}{3}
\nosection{Deployment of the HAWC gamma-ray observatory in Sierra Negra,
Mexico\\
{\footnotesize\sc Ibrahim Torres, Alberto Carrami\~{n}ana, Ruben Alfaro, Arturo
Iriarte}}
\setcounter{section}{0}
\setcounter{figure}{0}
\setcounter{table}{0}
\setcounter{equation}{0}
%%
% 33nd International Cosmic Ray Conference - 2013 - Rio de Janeiro, Brazil
% Template adapted from the 2011 ICRC template.
%
%\documentclass[a4paper]{article}
%
%\usepackage{icrc2013}

%The paper title
\title{Deployment of the HAWC gamma-ray observatory in Sierra Negra, Mexico}

%The short title to appear at the header of the pages.
\shorttitle{Deployment of the HAWC gamma-ray observatory in Sierra Negra, Mexico }

%All paper authors
\authors{
Ibrahim Torres$^{1}$, Alberto Carrami\~nana$^{1}$, Ruben Alfaro$^{2}$, 
Arturo Iriarte$^{3}$, for the HAWC Collaboration$^{4}$ }

%All the affiliations.
\afiliations{
$^1$Instituto Nacional de Astrof\'{\i}sica, \'Optica y Electr\'onica, Luis Enrique Erro 1,
Tonantzintla, Puebla 72840, M\'exico\\ 
$^2$Instituto de F\'{\i}sica, Universidad Nacional Aut\'onoma de M\'exico\\
$^3$Instituto de Astronom\'{\i}a, Universidad Nacional Aut\'onoma de M\'exico\\
$^4$For a complete author list, see the special section of these proceedings
}

%email address of the contact person
\email{ibrahim@inaoep.mx}

%The abstract.
\abstract{
The HAWC (High Altitude Water Cherenkov) gamma-ray observatory located at an altitude of 4100 meters in Sierra Negra, Puebla, M\'exico, is in its second phase of construction on a 22,500 m$^2$ platform. More than 70 water Cherenkov detectors of 4.5 m height and 7.3 m diameter have been deployed, each detector having three 8" PMT's and one 10" PMT in its interior, allowing efficient detection of Cherenkov radiation in the water due to secondary particles from atmospheric air showers. The +19$^\circ$N location of the site allows for a 2/3 coverage of the celestial sphere and the Galactic plane. We describe the construction of the HAWC instrument including the technique for constructing the water tanks, the development of the site infrastructure, and the water delivery, capture and filtration systems. }

%The keywords

%\hyphenpenalty=1000
%
%
%\begin{document}
\maketitle

%Begin a section.
\section*{Introduction}

The HAWC (High Altitude Water Cherenkov) observatory is a very high energy gamma-ray detector that will cover
the 100 GeV - 100 TeV energy range with a 1.8 ~sr instantaneous field of view during ten years of continuous 24h/day
operations. Its original specifications call for a water Cherenkov detector of 22,000 m$^2$ in area, located above 4000m 
of altitude, with a layer of 900 8" and 300 10" photomultiplier tubes placed below 4.5 m of water.  The construction of the observatory began in summer 2011 and the installation of the detector array in February 2012. 
The current status of the installation process is described here.

\section*{The HAWC site}
Volc\'an Sierra Negra, or Tliltepetl, is a 460,000 year old volcano hosting several scientific facilities benefiting from its 
high altitude and geographical location. The 4582 m summit was selected in February 1997 as the site of the Gran 
Telescopio Milim\'etrico Alfonso Serrano or Large Millimeter Telescope (LMT), now entering early science operations. 
Ten years later, the Northern base of Sierra Negra was chosen as the site of the High Altitude Water Cherenkov
(HAWC) gamma-ray observatory, a wide field of view instrument conceived to perform continuous monitoring of
the sky in the 100 GeV - 100 TeV energy range~\cite{icrc2007}.

\begin{figure*}[!t]
\centering
\includegraphics[width=0.9\textwidth]{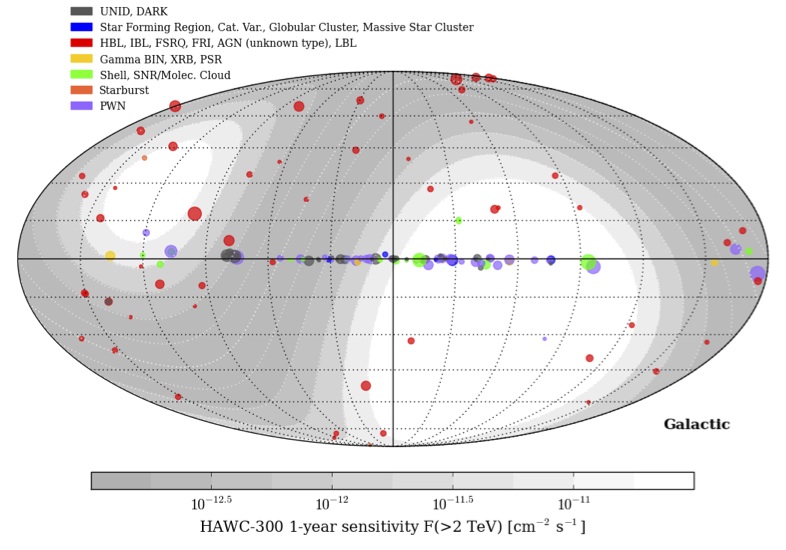}
\caption{Sky coverage of HAWC in Galactic coordinates.} \label{sky}
\end{figure*}

The geometrical center of the HAWC platform is located at 4097 m above sea level, which corresponds to an 
atmospheric depth of 638 g/cm$^2$, or 17.2 times the mean free path for e$^\pm$ pair production and
bremsstrahlung interactions in air at sufficiently high energies. This allows for secondary particles generated 
in electromagnetic cascades of primaries with energies as low as 10~GeV to be able to reach the ground detector.
The geographical latitude of the site is 18.99$^\circ$, which allows for observations of 8~sr, 
or 2/3 of the sky, corresponding to objects culminating within 45$^\circ$ of the local zenith. HAWC will have 
considerable daily exposure to sources like the Crab, Geminga, Mrk 421 and 501, and the Cygnus region.
HAWC will be able to reach even the Galactic Center as it culminates at 46$^\circ$ from the zenith
and stays above $50^\circ$ for two hours per day. Figure~\ref{sky} shows the coverage of HAWC in Galactic coordinates.
The geographical longitude of the HAWC site, $97.31^\circ$W, 
is similar to that of major observatories in Mexico, the United States and South America, facilitating multi-wavelength 
follow-up observations of HAWC triggered events.

The HAWC site is within the Parque Nacional Pico de Orizaba, hence federal property and its access has been
granted by permits from the Secretar\'{\i}a del Medio Ambiente y Recursos Naturales (SEMARNAT) based on the 
review and approval of the environmental impact declaration. All environmental permits were obtained by March 2009 
and the access road to the site was open in July of the same year. The site permit establishes four stages for the HAWC 
project: installation in three years; ten years of operations; three years for dismantlement and two years for environmental 
site recovery.

\section*{Site infrastructure, layout and platform}

The development of the HAWC project has benefited from the infrastructure of the Large Millimeter Telescope. The road, electricity and optical fiber were installed through a 700m derivation at a point just after the LMT security gate. This allowed to rapidly set the conditions for the creation of the HAWC platform and installation of the detector array.

HAWC has been designed to be an array of up to 300 individual water Cherenkov detectors (WCDs) covering an area of about
22,500{m$^2$. Each tank is a cylinder of 7.3m in diameter and 5m in height filled with water, isolated of environmental light and instrumented with four photomultiplier tubes (PMTs). Every WCD is controlled from the central Counting House where data are acquired. The array can be monitored remotely through a web interface. The layout of the tank array, shown in figure~\ref{layout}, considers the need of having an array as dense as possible  together with the needs of maintenance. Tanks are placed in 
joint pairs of rows, with enough clear space  to allow for maintenance and repairs of any tank. During 2011 an area of about 22,395~m$^2$, corresponding to the HAWC platform, was flattened to from an original South-North slope of about 8\% to 1\%, enough to allow the draining of precipitable water. The HAWC Utility Building (HUB) was installed East to the tank array platform (fig.~\ref{plataforma}). The HUB contains the water filtration system and the infrastructure required to test plastic bladders, prior to the installation of the individual WCDs.

 \begin{figure}[h]
  \centering
  \includegraphics[width=0.48\textwidth]{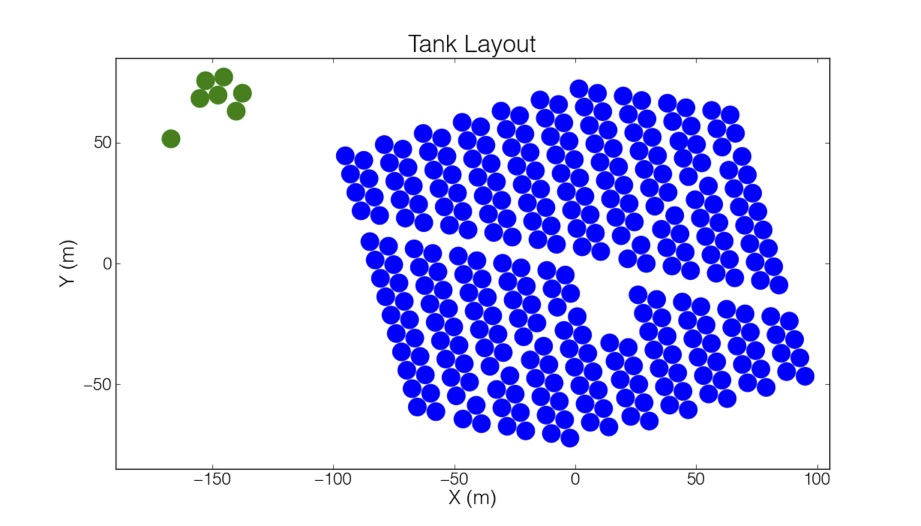}
  \caption{\small Layout of the HAWC observatory. The 300 blue clustered tanks constitute the detector array, with the gap corresponding with the location of the counting house. At the upper left is 
  the seven tanks of VAMOS engineering array, operated during 2011. The display in in local coordinates, inclined 16$^\circ$ relative to the geographical coordinates.}
 \label{layout}
 \end{figure}

 \begin{figure}[h]
  \centering
  \includegraphics[width=0.48\textwidth]{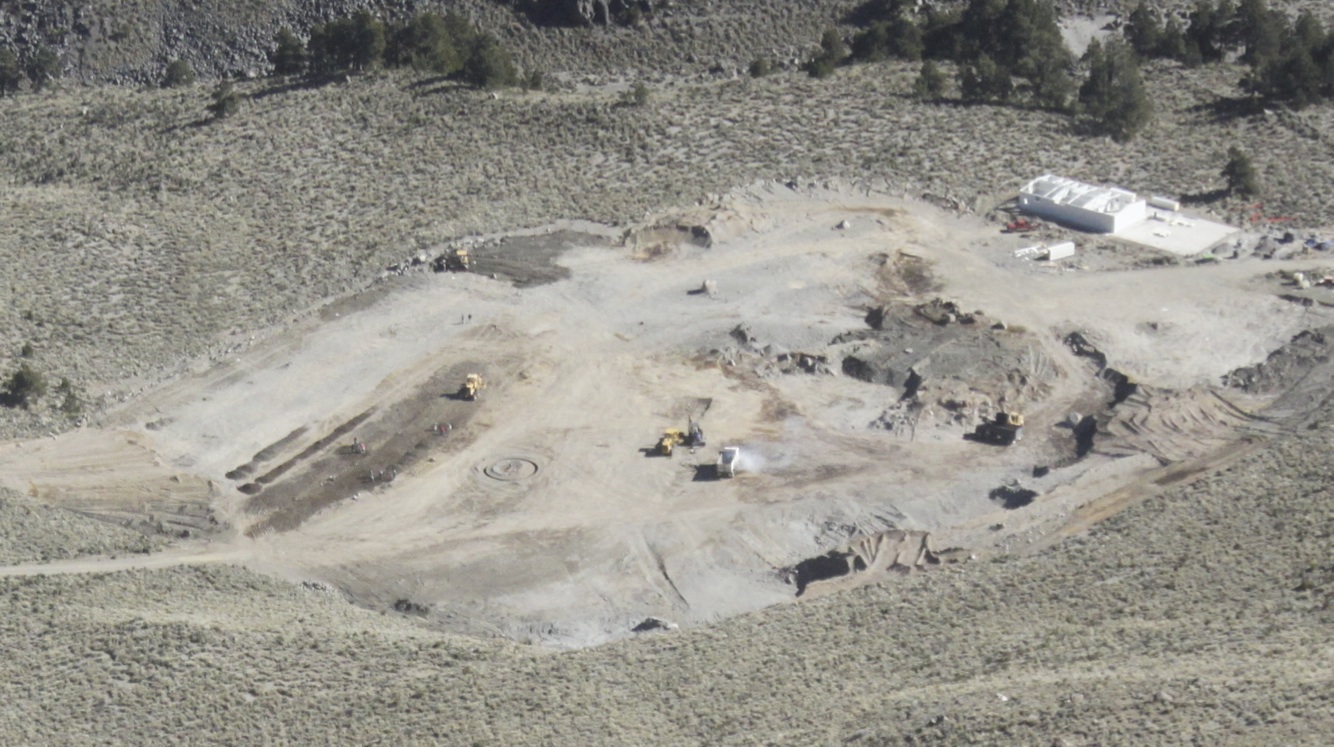}
  \caption{\small The HAWC platform in process of flattening, as October 2011. The HUB can be appreciated as the white construction at the right.}
 \label{plataforma}
 \end{figure}
 
 % \begin{figure*}[!t]
%  \centering
%  \includegraphics[width=\textwidth]{icrc2013-template-01}
%  \caption{Wide figure example. Conference webpage header.} \label{wide_fig}
% \end{figure*}

\section*{Design and construction of the water Cherenkov detectors}

The HAWC WCDs are steel tanks lined with a plastic bladder allowing to storage 188 000 litters of purified water. 
These cylindrical tanks are constructed from galvanized steel commonly used as silo structures. The tank walls vary in thickness with height to minimize the cost, since the cost of the tanks is driven 
by the commodity cost of steel. The stress calculations were done using API Standard 620 and the maximum allowable tensile stress for tension for bolted steel. The pre-fabricated tanks are delivered ready for assembly on a single pallet. All holes are pre-drilled and fasteners and other hardware are provided with the kit. The maximum tensile strength has also been calculated for the bolts/fasteners which are critical to the structure (including nuts and the torque). The overlap or joining of the individual panels is also critical for the integrity of the tank. The proper overlaps combined with number of bolts insure that the panels are working as designed and withholding the pressure as a ring. The current design avoids local stress points that could create deformations. A dome-roof structure reinforces the strength of the top ring and prevent of any water or snow accumulation. At the bottom part, the steel structures are buried 60 cm in the ground to provide a natural anchor for stability. The assembly of the panels is doing top-down, the main reason for this is to deploy the roof at the ground level minimizing unsafe work practices.
 
The procedure starts with a surveyed stake as a center reference, around it a circular trench is excavated. Eight foundation blocks are accurately positioned and leveled in the trench. They serve to set the height of the finished tank and act as a level base for the jack stands used during construction. Several tubular steel spanners are placed over the trench. The top ring of the tank is assembled on these spanners over the trench. The roof is then attached to the top ring. An access area, large enough to readily accommodate a person, is left open. Four jack stands, positioned on the inside of the tank, are bolted to the top ring. The feet of the stands are positioned on the foundation blocks. A mechanical jack is attached to the jack stand. The top ring and roof are jacked up and the stands are securely pinned. The second ring is assembled and bolted on below the first. The stands are lowered and attached to the second ring. The two rings and roof are jacked up and the stands securely pinned. The third ring is assembled and bolted together below the second. This cycle is repeated for the fourth and fifth rings. When the fifth ring is completed, the whole structure is slowly lowered into the trench. The trench is filled in and the inside of the tank is leveled.

\section*{Current status}

At the time of writing, more than 100 WCDs have been assembled; of these 97 have been water filled and 86 instrumented with PMTs. This will allow HAWC to enter in science operations mode from August 2013, with the final installation goal of completing 300 WCDs at the end of 2014 and operate the array for five to ten years. HAWC has been in a pre-operational mode since September 2012, acquiring enough data to make a high significance detection of the Moon shadow on the cosmic-ray background and the anisotropy in the arrival direction of cosmic-rays~\cite{hawc-moon,anisotropy}. The study of solar events together with searches for gamma-ray bursts and flares from active galactic nuclei are now on-going.

\begin{figure*}
\centering
\includegraphics[width=\textwidth]{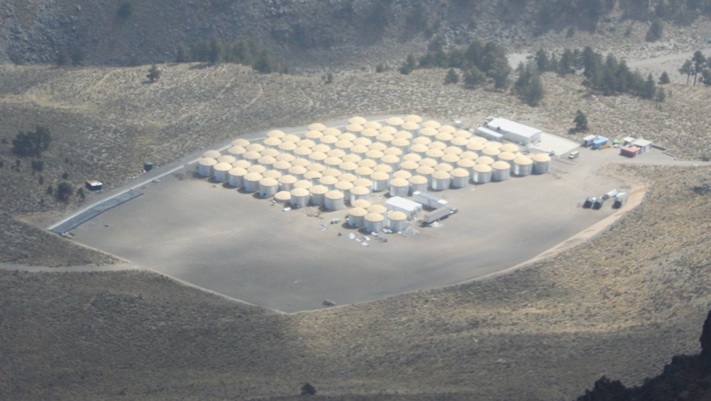}
\caption{View of HAWC with over 100 tanks (May 2013).} \label{hawc-may}
\end{figure*}

%\section*{Conclusions}

% A list of references should be placed at the end of the paper. See the below example. The citation to the references 
% can be written as \cite{bib:lattes} or \cite{bib:lattes,bib:schoenberg}.

\section*{Acknowledgments}

The HAWC is a wide M\'exico - United States scientific collaboration comprising the following institutions: \\
{\bf M\'exico:} Instituto Nacional de Astrof\'{\i}sica, \'Optica y Electr\'onica; Universidad Nacional Aut\'onoma de M\'exico (Instituto de Astronom\'{\i}a; Instituto de F\'{\i}sica; Instituto de Ciencias Nucleares; Instituto de Geof\'{\i}sica); Benem\'erita Universidad Aut\'onoma de Puebla; Universidad Aut\'onoma de Chiapas; Universidad Michoacana de San Nicol\'as de Hidalgo; Universidad de Guadalajara; Universidad de Guanajuato; Instituto Polit\'ecnico Nacional y Centro de Investigaci\'on y Estudios Avanzados del IPN; Universidad Aut\'onoma del Estado de Pachuca; and Universidad Polit\'ecnica de Pachuca.\\
{\bf United States:} University of Maryland; Los Alamos National Laboratory; Colorado State University; George Mason University; Georgia Institute of Technology; Michigan State University; Michigan Technological University; NASA Goddard Space Flight Center; Pennsylvania State University; The Ohio State University; University of Alabama; University of California at Santa Cruz; University of California Irvine; University of New Hampshire; University of New Mexico; University of Utah; and University of Wisconsin.\\

The HAWC $\gamma$-ray observatory is supported by the National Science
Foundation and the Department of Energy of the United States, and by the
Consejo Nacional de Ciencia y Tecnolog\'{\i}a in M\'exico.

%\end{document}

% Timing calibration
\newpage
\setcounter{section}{4}
\nosection{Timing Calibration of the HAWC Observatory\\
{\footnotesize\sc Hugo~A. Ayala Solares, Hao Zhou, Chiumun Michelle Hui, Petra
H\"{u}ntemeyer}}
\setcounter{section}{0}
\setcounter{figure}{0}
\setcounter{table}{0}
\setcounter{equation}{0}
%%
% 33nd International Cosmic Ray Conference - 2013 - Rio de Janeiro, Brazil
% Template adapted from the 2011 ICRC template.
%
%\documentclass[a4paper]{article}
%
%\usepackage{icrc2013}
%\usepackage{textcomp}
%\usepackage{lineno}
%\linenumbers
%The paper title
\title{Timing Calibration of the HAWC Observatory}

%The short title to appear at the header of the pages.
\shorttitle{HAWC Timing Calibration}

%All paper authors
\authors{
H. A. Ayala Solares$^{1}$,
H. Zhou$^{1}$,
C. M. Hui $^{1}$,
P. H{\"u}ntemeyer $^{1}$,
for the HAWC Collaboration$^{2}$.
}

%All the affiliations.
\afiliations{
$^1$ Michigan Technological University \\
$^2$ For a complete author list, see the special section of these proceedings \\
}

%email address of the contact person
\email{hayalaso@mtu.edu, hzhou1@mtu.edu, cmhui@mtu.edu, petra@mtu.edu}

%The abstract.
\abstract{The High Altitude Water Cherenkov (HAWC) Observatory is a gamma-ray experiment being built in Mexico. It will be an array consisting of 300 water Cherenkov detectors (WCDs). Four photomultiplier tubes (PMTs) will be deployed on the bottom of each WCD to detect the Cherenkov light produced by the secondary particles in air showers caused by the interaction of cosmic particles and high-energy gamma rays with the atmosphere. The relative times between the PMT channels are crucial for reconstructing the direction of the primary particle. The response time of a PMT and electronics depends on the light intensity striking on the PMT. A laser calibration system was designed to accurately measure the relative timing among the PMT channels. Laser pulses with varying intensities are sent to each WCD through optical splitters, switches, and fibers. The time between the laser shot and the PMT signal is recorded to correct for the dependence on the light intensity. A time residual study is also performed to improve the angular reconstruction. The time residual is the difference between a fitted air shower front and the PMT readout time. This systematic time offset is then accounted for in an iterative shower reconstruction procedure to improve the determination of the incoming direction of the primary particle. A partial array of 30 WCDs began operation in Fall 2012. In this contribution, the first results of the timing calibration curves and time residual studies are presented.
}

%The keywords

%\begin{document}
\maketitle

%Begin a section. 
\section*{Introduction}

The High Altitude Water Cherenkov (HAWC) Observatory is a gamma-ray experiment being built in Mexico. It is located at 4100 m above sea level on the flanks of the volcano Sierra Negra in the state of Puebla, Mexico. It will be an array consisting of 300 water Cherenkov Detectors (WCDs) covering an area of $\sim$22,000 $m^2$. Each tank is 7.3 m in diameter and  4.5 m deep and it is filled with $\sim$200,000 $L$ of purified water. Four photomultiplier tubes (PMTs) will be deployed on the bottom of each WCD to detect the Cherenkov light produced by the secondary particles in air showers caused by the interaction of cosmic particles and high-energy gamma rays with the atmosphere.  With a duty cycle of $>$90$\%$ and an instantaneous field of view of 2 sr, HAWC will be the most sensitive detector in the gamma-ray energy range of 0.1 TeV and 100 TeV. (See \cite{bib:Mostafa} for more details).

Reconstructing the direction of the primary particles does not require the exact absolute time when a PMT is hit by the front of an extensive air shower, but the accurate relative times between the PMT channels are crucial. A timing calibration on each PMT in the array is essential to reach the goal of an angular resolution 0.35 \textdegree - 0.1 \textdegree \cite{bib:Calibration}. In this proceeding a description of different timing calibration measurements will be given. 

The first of these calibrations is the slewing time, namely, the response time of a PMT and electronics. To obtain the real PMT hit time for angular reconstruction, the slewing time, which depends on the light intensity striking on the PMT, needs to be subtracted from the measured PMT hit time. A laser calibration system was designed to accurately measure the slewing time and correct it from the air shower data. A description of the laser calibration system and the first results on a subset of 30 WCDs of the array will be given in section 2.

The second time measurement is a time residual defined as the difference between a fitted air shower front expected time and the PMT readout time.  This systematic time offset is then accounted for in an iterative shower reconstruction procedure to improve the determination of the incoming direction of the primary particle. In section 3 we will talk about these time residuals and we will show some of the results that we already have with a subset of 30 WCDs of the array that are already installed.

In section 4 we will give our conclusions.

\section*{Slewing Time}

In HAWC, two thresholds (low and high thresholds) are used to sample both small and large PMT pulses accurately. Each time a PMT pulse crosses any of the two thresholds, a time stamp is given and recorded as an edge by a time-to-digital converter (TDC) with a precision of 0.1 ns. The width and ultimately the size of the pulse can be inferred from the time-over-threshold (ToT) for the low and/or high thresholds. 

The time of the first edge (i.e. the time of the pulse crossing the low threshold) is considered as the time of the pulse. However, the rising edge of a larger pulse will cross the thresholds earlier, yielding an earlier signal and a shorter response time. The slewing time, which is the gap between the PMT hit time and the first edge recorded by the TDC, depends on the size of the pulse thus needs to be corrected in order to obtain the real hit time for resolution \cite{bib:HaoSlewing}.

\vspace*{0.5cm}

\subsection*{Laser Calibration System}

A laser calibration system was designed to measure the dependence of the slewing time on the pulse size (ToT). A series of laser pulses are attenuated by a set of different neutral density filters to provide a wide range of laser intensities over 6 orders of magnitude. The pulses are delivered into 300 WCDs through a set of optical splitters, switches and fibers \cite{bib:Calibration}. In addition to the light paths to the WCDs, one of the light paths will send laser to a Thorlab photo sensor to produce a start signal. The time from the start signal to the pulse crossing the low/high threshold in each PMT channel is measured as slewing time for low/high threshold.

\subsection*{Results and Discussion}

The laser calibration system was installed at the site in last December and the first tests and results on a subset of 30 WCDs have been obtained. Figure \ref{E18} shows an example of slewing curves of four PMTs in WCD E18.  The central PMT E18C is a high quantum efficiency 10-inch PMT while the other three are 8-inch PMTs. The upper four curves (dashed) are slewing curves for high threshold, and the lower four curves (solid) are slewing curves for low threshold. The distribution of slewing time vs. ToT is binned by ToT with 10 ns bin width, and the gaussian mean and sigma of slewing time in each ToT bin are the data point and error bar in the plot. A dedicated function (equation \ref{eq:6par}) is fit to the slewing curves for both low and high thresholds.

\begin{equation}
\ Slewing Time = e^{\frac{-ToT-p0}{p1}} - e^{\frac{ToT-p2}{p3}} + p4 - p5 * ToT
\label{eq:6par}
\end{equation}

\begin{figure}[t]
  \centering
  \includegraphics[width=0.4\textwidth]{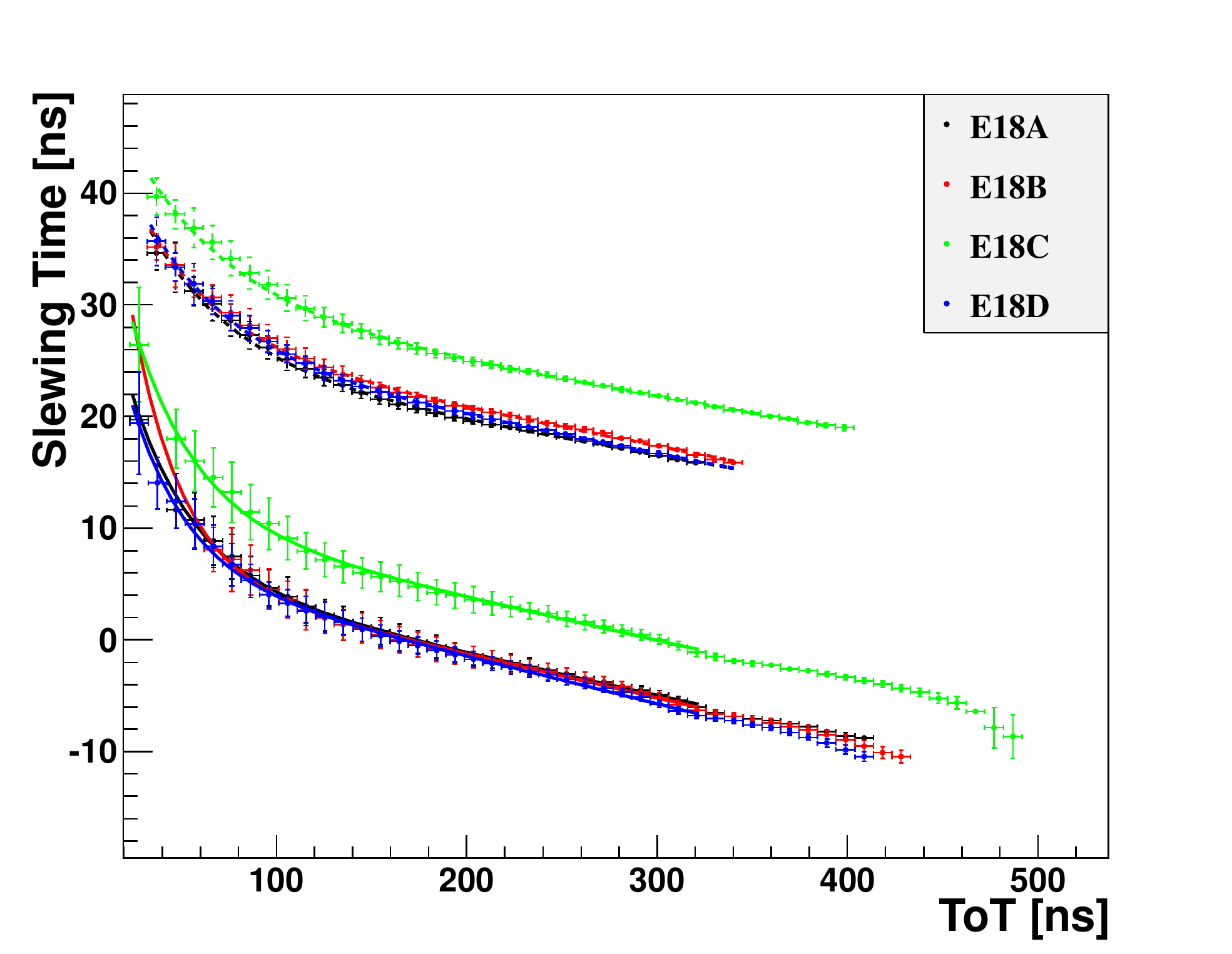}
  \caption{Example of slewing curves of four PMTs in WCD E18. E18C is HQE 10-inch PMT and the other three are 8-inch PMT. Upper four curves (dashed) - high threshold. Lower four curves (solid) - low threshold. The x axis is ToT (ns) and the y axis is the time (ns) between the start signal to the PMT signal crossing the threshold (an offset of 1782 ns subtracted).}
  \label{E18}
\end{figure}

Figure \ref{Average} shows the average slewing curves in 30 WCDs for 8-inch and 10-inch PMTs, respectively. The lighter curves on both side of the average curves show the one sigma error of the curves for different PMTs and the lightest curves show the minimum and maximum values among the curves. To compare the shape of the slewing curves for different PMTs, the time offsets, which can be corrected by the time residuals (see section 3), are removed by shifting the curves. The variation among the curves is less than a few ns at any ToT value after shifting the curves to a reference point at 200 ns of high ToT.

\begin{figure}[t]
  \centering
  \includegraphics[width=0.4\textwidth]{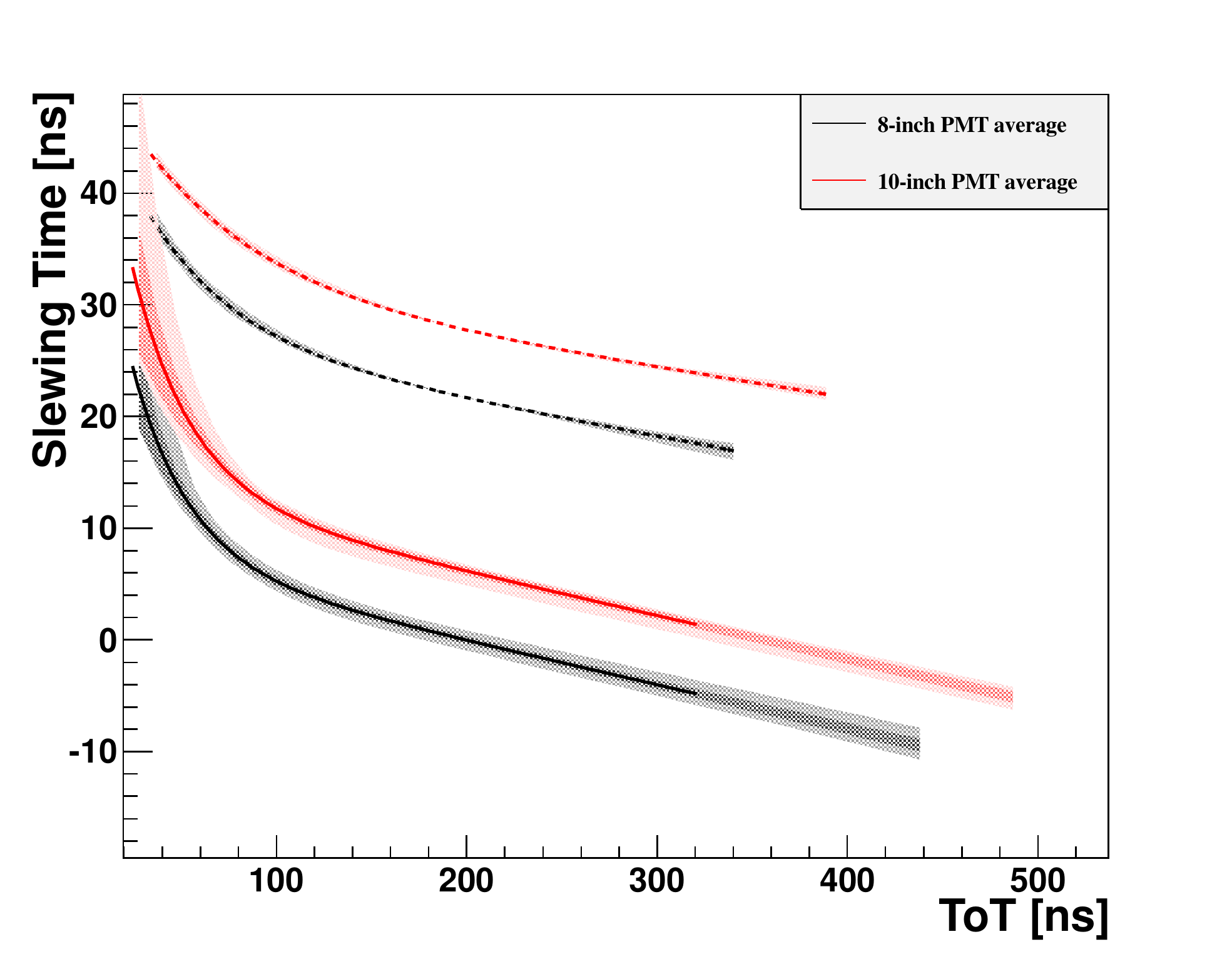}
  \caption{Average, one-sigma, min/max slewing curves for 8-inch and 10-inch PMTs. The individual curves are shifted to a reference point at 200 ns of high ToT.}
  \label{Average}
\end{figure}

The performance of the slewing calibration applying to the air shower data will be shown in section 3.1.1.

\section*{Time offset from the air shower front: Time Residuals}
\subsection*{Time Residual Definition}
In order to understand the time offset from the air shower front we need to describe how HAWC does the reconstruction process.
The reconstruction process has as one of its objectives to obtain the position of the shower core as well as  the direction of the primary gamma ray. With HAWC, the procedure is as follows:

First the core position of the shower is located. The core position of an extensive air shower is defined as the point in the ground where the gamma ray would hit if it did not interact with the atmosphere. The core contains the highest energy particles in the shower and this will be detected by the charge in the PMTs \cite{bib:abdo}.

The second part of the reconstruction process is to determine the direction of the incoming gamma ray. The procedure consists on finding the air shower front. In general the shower front is approximately spheric. In order to get the direction, the shower front is transformed into a plane, so that the normal of this plane points where the primary particle comes from \cite{bib:abdo}.
To modify the shower front, we use the location of the PMTs with respect to the core position and the charge in the PMTs (See \cite{bib:lauer}).

The time residuals (TR) are a systematic time offset which is defined as the difference between the fitted air shower front expected time and the PMT readout time. A diagram showing the curvature correction and the time offset is in figure \ref{diagram}. This systematic time offset is then accounted for in an iterative shower reconstruction procedure to improve the determination of the incoming direction of the primary particle.

\begin{figure}[t]
  \centering
  \includegraphics[width=0.4\textwidth]{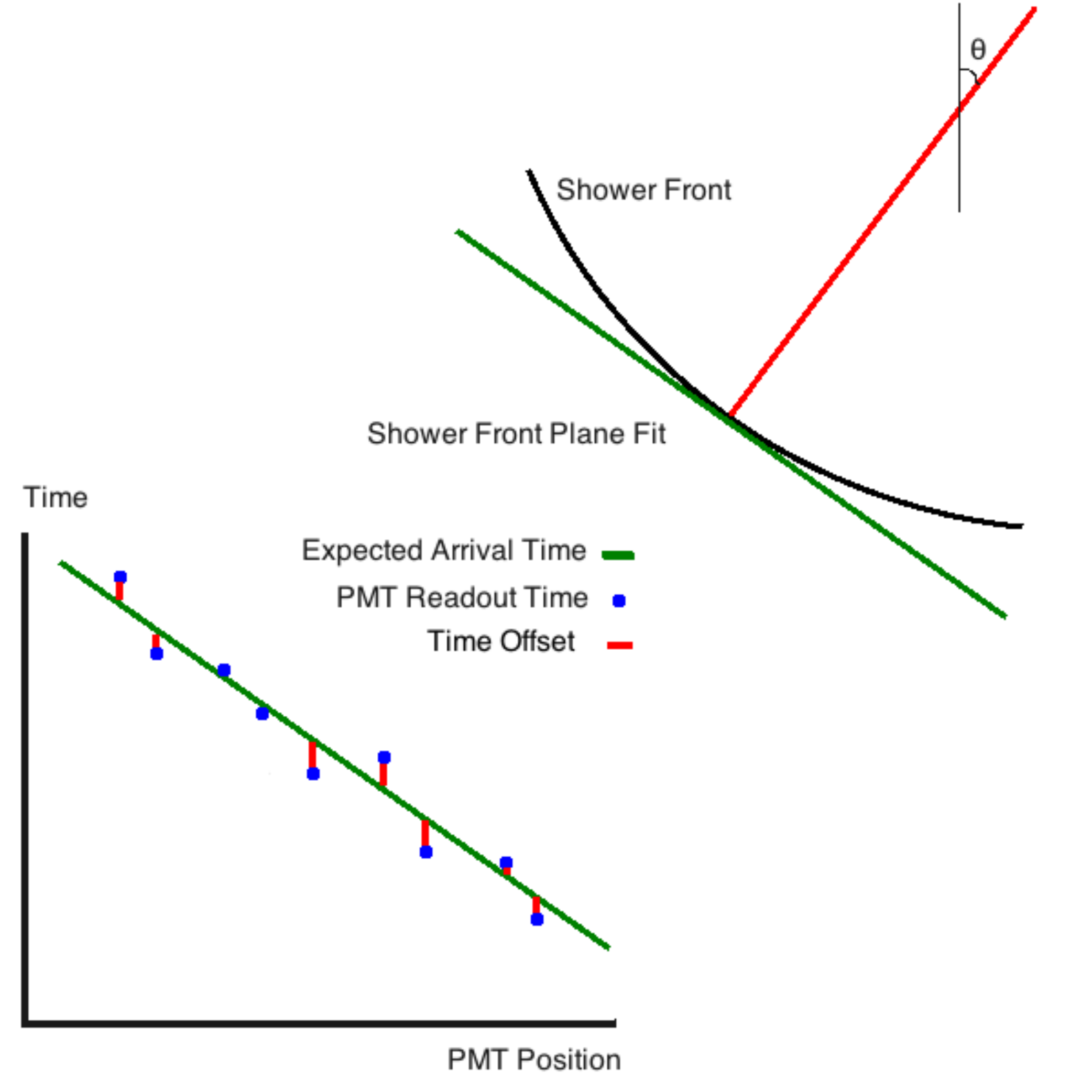}
  \caption{The black line represents the true shower front of an extensive air shower. The green line represents the plane shower front after the curvature correction. This is use for the expected arrival time, which is then compared to the readout times of the PMTs}
  \label{diagram}
 \end{figure}
 
\begin{equation}
 TR = PMTTime - ExpectedTime
\end{equation}

A histogram for each PMT is obtained from the data. An example is shown in figure \ref{fit}. If the peak of the histogram is positive, then the hits were late for that channel.  If the peak is negative, then hits were early. During the iterative procedure, we need to make the hits earlier in the first case and later in the second case and to do so we just add the time residuals to the calibrated time. In the next subsection we will talk more about the fitting procedure and the iteration process. \cite{bib:lansdell}

\subsection*{Results and Discussion}
In order to get the histograms for each PMT, we require certain cuts in the data:
\begin{itemize}
\item The events need to succeed the shower plane fit procedure so that we can have the expected time from the shower front
\item More than 40 hits in an event to use larger showers which will trigger more PMTs. 
\item More than 10 Photo-electrons in a hit. This will decouple the dependence of the electronic response time described in section 2 to the time residuals.
\end{itemize}
The peak of the histograms are obtained by fitting a gaussian function (Figure \ref{fit}).
 
 \begin{figure}[h!]
  \centering
  \includegraphics[width=0.4\textwidth]{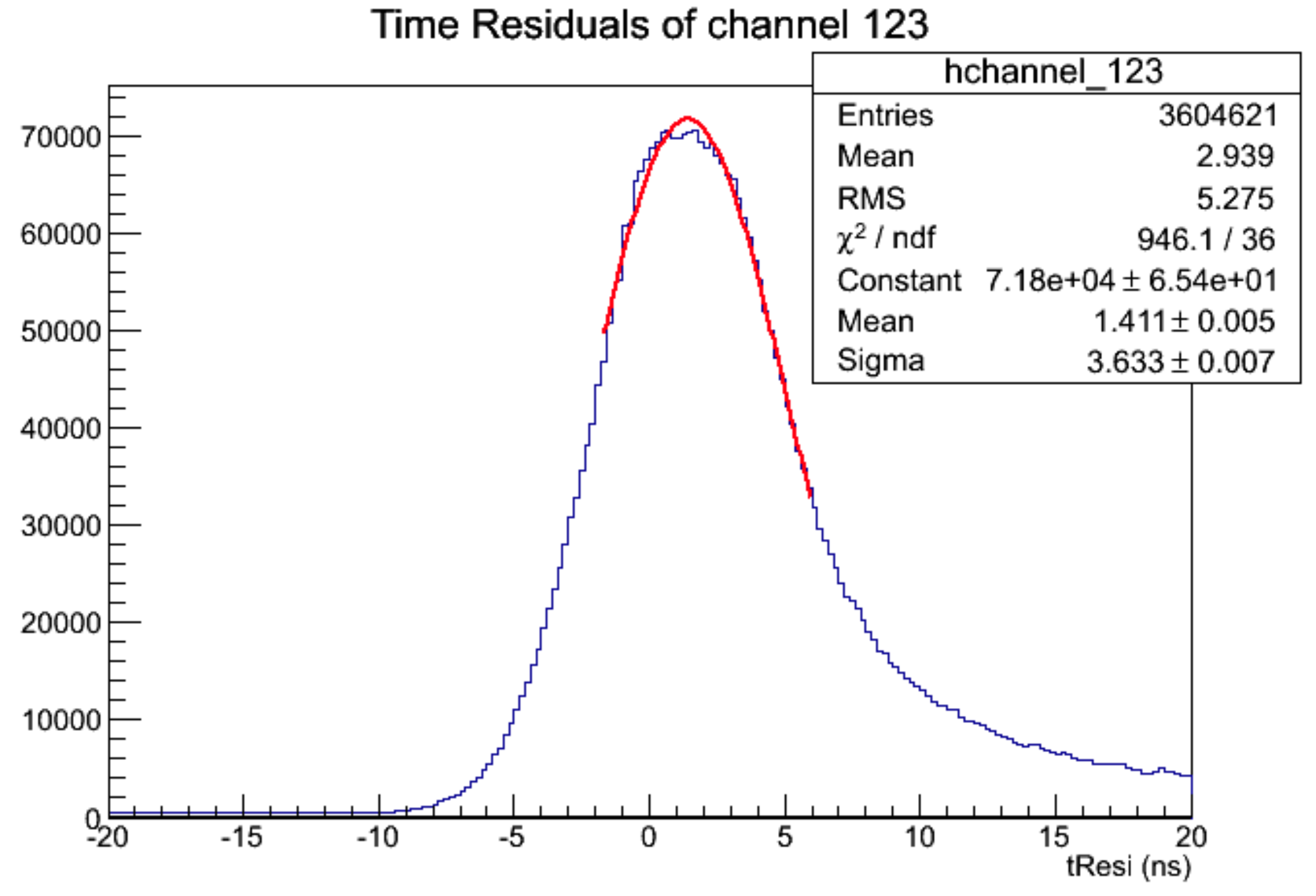}
  \caption{A fit of a histogram of a specific channel.}
  \label{fit}
 \end{figure}

For the iterative process, there is no specific number of iterations. We can see in figures \ref{nocorr} and \ref{corr}  that 3 iterations do the work of reducing the time residuals. 

 \begin{figure}[h!]
  \centering
  \includegraphics[width=0.4\textwidth]{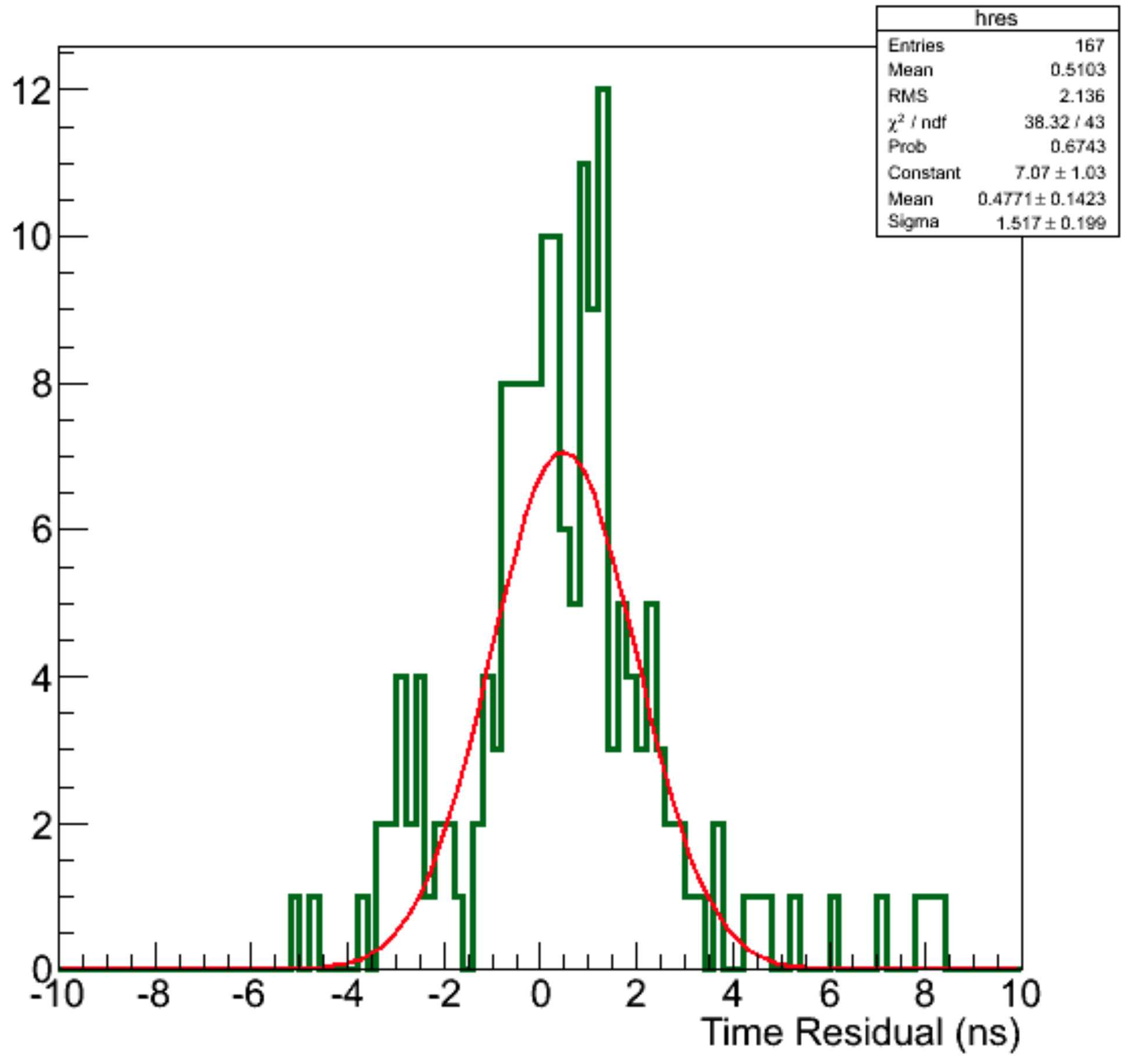}
  \caption{Mean time residuals before the correction. A gaussian fit is performed to give an idea of the spread of the mean time residuals.}
  \label{nocorr}
 \end{figure}
 
 \begin{figure}[h!]
  \centering
   \includegraphics[width=0.4\textwidth]{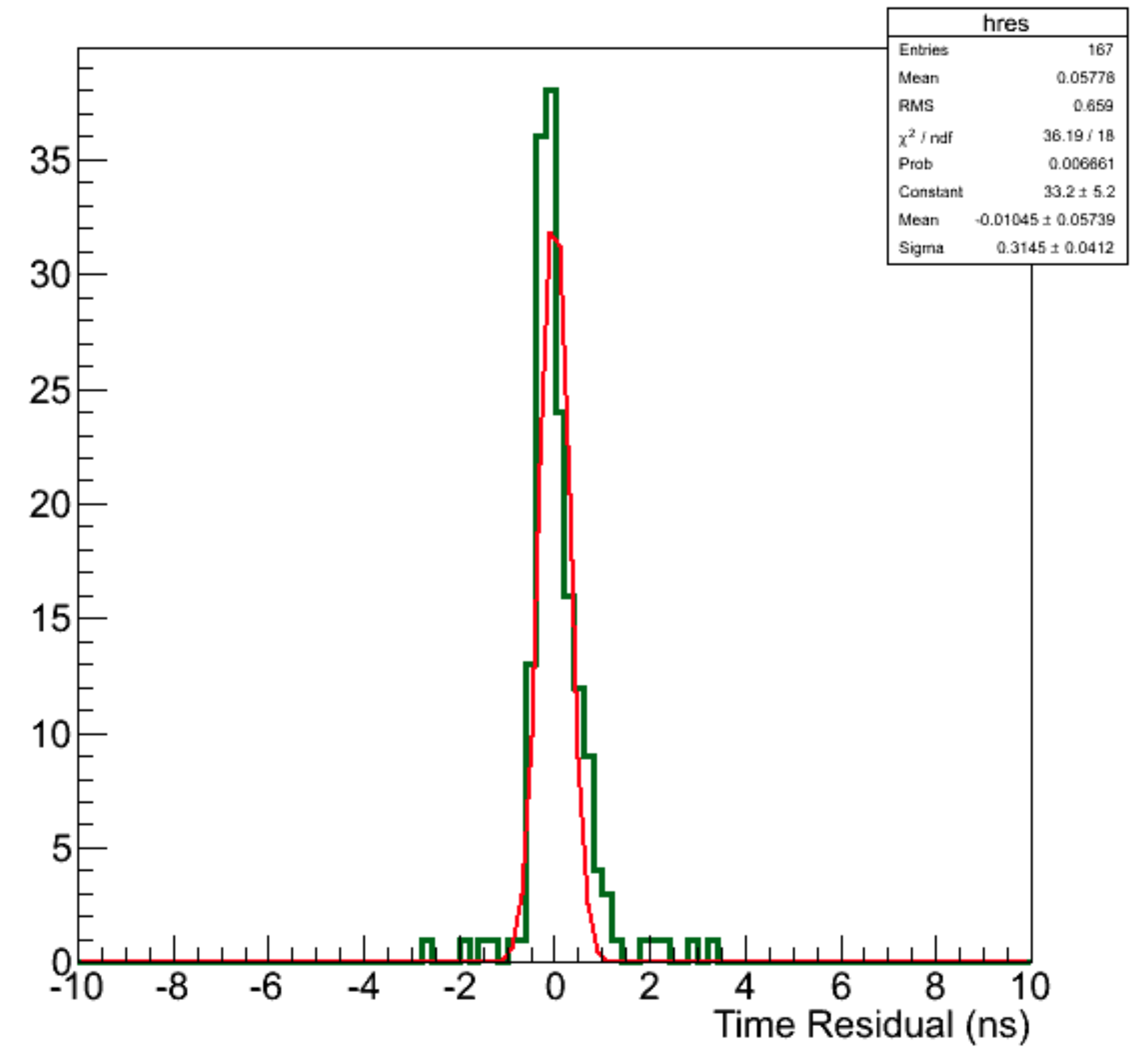}
  \caption{Mean time residuals after a 3 iteration process. A gaussian fit is performed to see how the spread is of the mean time residuals is reduced after the iteration process.}
  \label{corr}
 \end{figure}

Also the time residuals after the iteration process should have an accuracy of $\sim$0.1 ns. An example of a fit after the applying the time residuals is shown in figure \ref{fit2}.
  \begin{figure}[h!]
  \centering
  \includegraphics[width=0.4\textwidth]{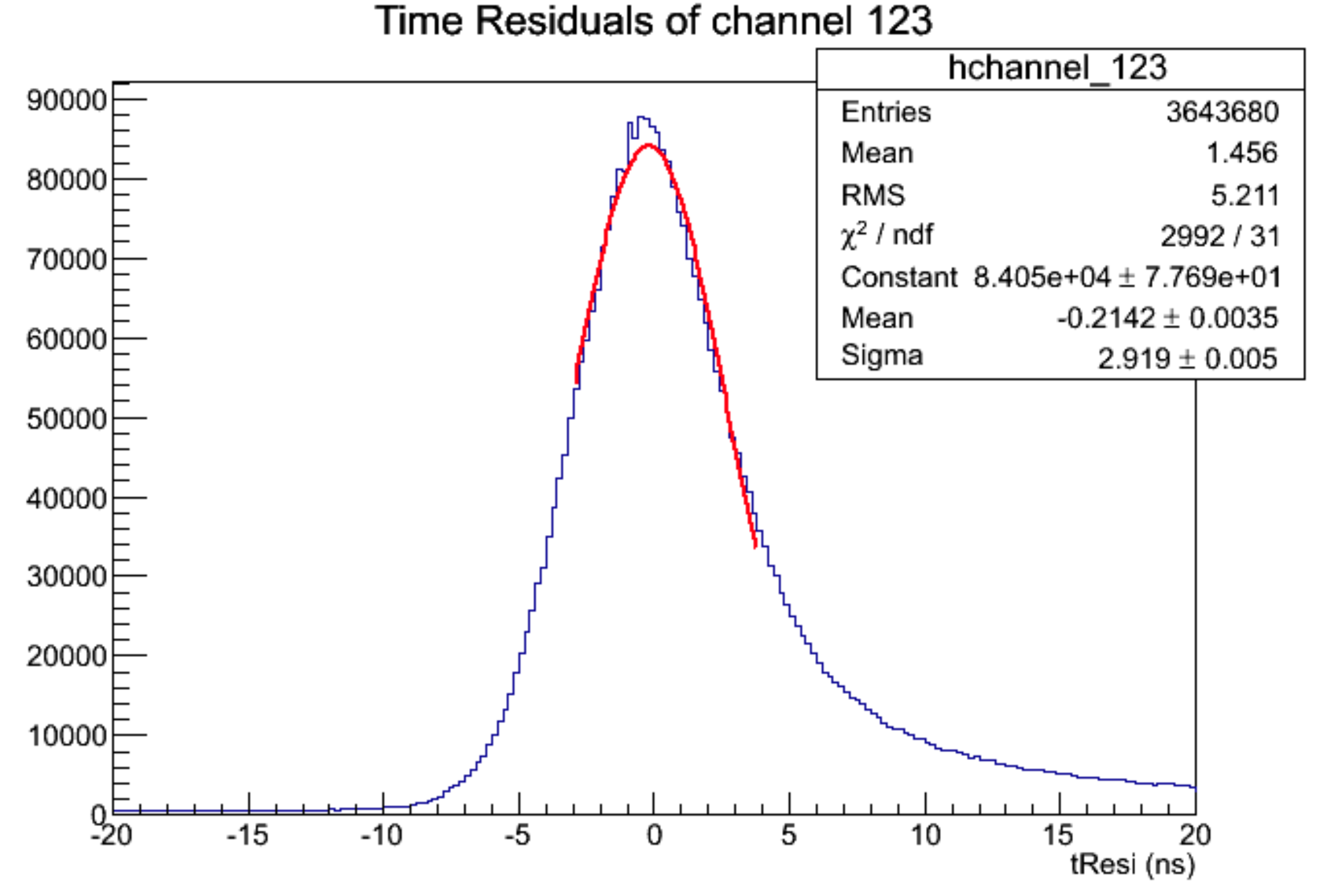}
  \caption{Fit of time residual histogram after the correction. The accuracy of the mean time residual needs to be around $\sim$0.1 ns which is the resolution of the TDCs.}
  \label{fit2}
 \end{figure}
The performance of the data analysis using the timing calibrations can be seen in \cite{bib:lauer}, specifically figure 2c of that proceeding.
\newpage
\subsection*{Timing Calibration on Air Shower Data}

Both slewing calibration and time residual correction are applied to the PMT readout times before shower reconstruction. Figure \ref{SlewingCorr} shows the remaining time residual vs. low ToT after shower reconstruction. Without timing calibration, time residuals are highly dependent on ToTs due to slewing effect. Large error bars are contributed from offsets between PMT channels and from shower thickness.

With slewing calibration, the dependence of time residuals on ToTs has been mostly removed, resulting in a flat curve. The offsets between PMT channels are removed by time residual correction, resulting in smaller error bars.

\begin{figure}[h!]
  \centering
  \includegraphics[width=0.4\textwidth]{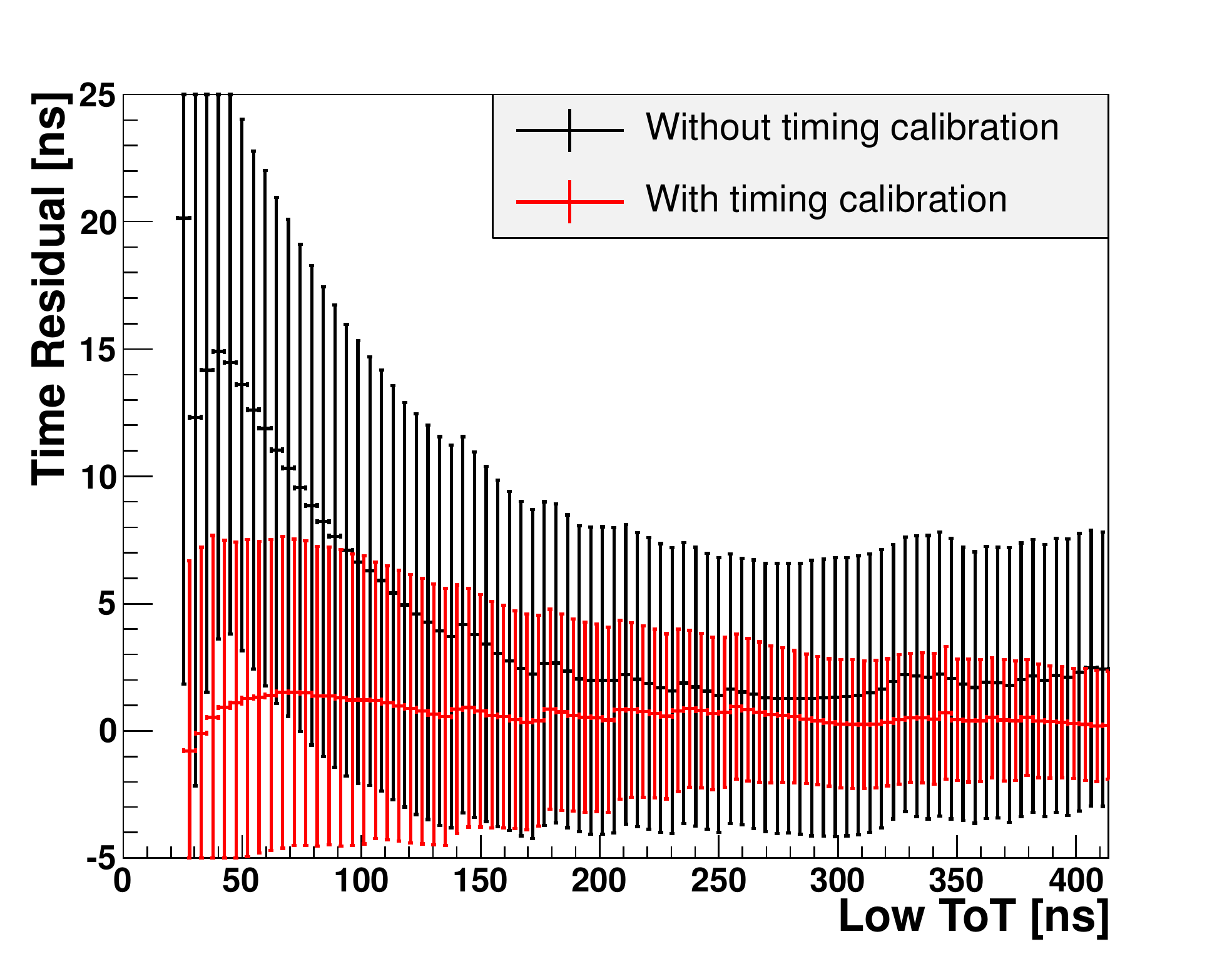}
  \caption{Correlation between the time residual and low ToT without (black) and with (red) timing calibration. The x axis is low ToT (ns) and the y axis is the time residual (ns). Data point - gaussian mean. Error bar - one sigma.}
  \label{SlewingCorr}
\end{figure}

\section*{Conclusions}
The timing calibration of HAWC is one of the crucial tasks since the relative times between the PMTs are used for the angular reconstruction of the air shower events. A description and task of the different times associated with the timing calibration was given in this proceeding.

With slewing calibration, the dependence of the slewing time on the pulse size (ToT) is significantly reduced, resulting in more accurate PMT hit times and more reliable angular reconstruction. 

Taking into consideration the systematic offset from the difference of the fitted shower front and the readout time of the PMTs is important since this will improve the reconstruction of the direction of the extensive air shower. To do so, an accuracy of $\sim$0.1 ns is necessary. As it was shown, an iterative procedure is used in order to get the accuracy mentioned for each PMT.

\section*{Acknowledgments}

We acknowledge the support from: US National Science Foundation (NSF); US
Department of Energy Office of High-Energy Physics; The Laboratory Directed
Research and Development (LDRD) program of Los Alamos National Laboratory;
Consejo Nacional de Ciencia y Tecnolog\'{\i}a (CONACyT), M\'exico; Red de
F\'{\i}sica de Altas Energ\'{\i}as, M\'exico; DGAPA-UNAM, M\'exico; and the
University of Wisconsin Alumni Research Foundation.

\clearpage

%\end{document}

% Calibration performance
\newpage
\setcounter{section}{5}
\nosection{Calibration and Reconstruction Performance of the HAWC Observatory\\
{\footnotesize\sc Robert Lauer}}
\setcounter{section}{0}
\setcounter{figure}{0}
\setcounter{table}{0}
\setcounter{equation}{0}
%%
% 33nd International Cosmic Ray Conference - 2013 - Rio de Janeiro, Brazil
% Template adapted from the 2011 ICRC template.
%
%\documentclass[a4paper]{article}
%
%\usepackage{icrc2013}
%\usepackage[english]{babel}
%\hyphenpenalty=1000

%The paper title
\title{Calibration and Reconstruction Performance of the HAWC Observatory}

%The short title to appear at the header of the pages.
\shorttitle{Calibration of the HAWC Observatory}

%All paper authors
\authors{
Robert J. Lauer$^{1}$
for the HAWC Collaboration.
}

%All the affiliations.
\afiliations{
$^1$ Department of Physics \& Astronomy, University of New Mexico, Albuquerque\\
}

%email address of the contact person
\email{rlauer@phys.unm.edu}

%The abstract.
\abstract{
The High Altitude Water Cherenkov (HAWC) experiment is being built at an altitude of 4100~m at Sierra
Negra volcano near Puebla, Mexico, to serve as an observatory for gamma-rays with energies between 50 GeV and 100 TeV.
Upon completion, the array will consist of 300 water Cherenkov detectors (WCDs) each equipped with four photo-multiplier
tubes (PMTs) to detect Cherenkov light from air showers passing through the experiment. For optimal reconstruction of
the direction of the shower primaries, in particular to map gamma-ray events, the relative timing of each PMT must be
calibrated with nanosecond (ns) precision over the full dynamic range of the PMTs. This is achieved by sending short
(300~picoseconds) laser light pulses through a network of optical fibers into every detector.
%To guarantee time stability of the laser pulses, the light transit times in the fibers are monitored. 
As the laser calibration system provides intensities from $<0.1$~photo electrons (PEs) to $>1000$~PEs, the charge
calibration of all PMTs is also obtained and further improves the angular resolution of gamma-ray event
directions.
In this contribution, we present the design and performance of the laser calibration system. The improvement of
air shower reconstruction due to calibration is verified for an analysis of the shadow that the moon produces in the
flux
of cosmic rays, based on data collected with a partial array of 30 WCDs.}

%The keywords

%\begin{document}
\maketitle

%Begin a section.
\section*{The HAWC Observatory}

Gamma-ray astronomy has become a field of rapid progress, with imaging air Cherenkov telescopes and
satellites detectors continuing to probe the sky at MeV to TeV energies. In a complimentary approach,
the Milagro Observatory~\cite{bib:Milagro} has proven that the water Cherenkov technique allows a ground based
TeV gamma-ray detector to operate with a high duty cycle and a wide field of view. The HAWC Observatory is being
built
as Milagro's successor, based on the same principle but surpassing Milagro's sensitivity by a factor of 15,
see~\cite{bib:HAWCgrb},\cite{bib:ICRC13general},\cite{bib:ICRC13sensi} for details. The higher altitude of 4,100~m above
sea level at the
HAWC site on the Sierra
Negra volcano near Puebla, Mexico, improves the low energy response compared to Milagro, widening it to approximately
50~GeV to 100~TeV.
HAWC's modular design comprises a total of 300 WCDs each equipped with three 8-inch PMTs and one
central 10-inch PMT.
These detectors record the times and multiplicities of Cherenkov photons from high energy particles
passing through the array and are used to reconstruct directions and energies of air shower primaries.
HAWC has a duty cycle of $>90$~\% due to its fully enclosed WCD design and thus serves as a powerful instrument
to
monitor and survey the TeV sky. In September 2012, the first 30 WCDs of HAWC became operational, making it possible to
start data taking while construction continues.
The experiment will transition smoothly to full scientific operation
with 100 WCDs in August 2013 and 300 WCDs in the summer of 2014.

The main data acquisition (DAQ) system is composed of time-to-digital converters (TDCs) with one channel for
each PMT.  When a photon induced pulse in a PMT crosses the low hardware threshold ($\sim0.35$~PE) or a high threshold
($\sim8$~PE), hit times get recorded as TDC counts ($10.24$~counts~$=1$~ns) relative to a trigger time. The
time-over-threshold (ToT) of a pulse for the low or high threshold can be converted into charge values, as described in
section~\ref{charge}.

The calibrated charges of all PMT pulses that are part of an air shower event are used as weights to locate the
core of the shower with a Gaussian fit of the charge distribution. This core fit serves as input to a second algorithm,
the angle fit, that calculates the curved shower front based on the photon arrival times and thus yields a direction
result for the event.
The ability to reconstruct this direction of an air shower primary depends crucially on a precise time record for each
PMT pulse. Statistical fluctuations introduce an irreducible spread of photon hit times on the order of a few ns
around the passing shower front, thereby defining the goal to calibrate the pulse timing to at least this accuracy. The
following sections discuss how these charge and timing calibrations are achieved in HAWC with a laser system.

\section*{The Laser Calibration System}

The light source components of the HAWC calibration system were installed between December 2012 and March 2013 in
the electronics facility in the center of the array. The calibration system relies on short laser pulses with a
width of 300~ps, created with a {\it Teem
Photonics} laser at a wavelength of 532~nanometers. An optical splitter cube located next to
the laser creates a separate {\it return light path}, while the primary or {\it light-to-tanks} path, leads
through a series of three filter wheels.
Each of these wheels carries six neutral density filter disks of different
absorption strength that can be cycled to attenuate the laser light. Aside from an open and an opaque setting, the range
of filter wheel combinations covers optical depths from $0.2$ to $8.0$.

The primary light-to-tanks path after the filter wheels fans out into optical fibers via {\it Dicon} switches that can
distribute the light into any of 150 separate channels, with up to 10 channels being illuminated simultaneously. At an
optical patch panel each channel is connected to a splitter that passes the same laser pulses into a pair of long, 170~m
optical fibers going out to a connection box near a pair of WCDs.
Light from the two outputs is then directed through short fibers into the two WCDs and exits through a Teflon diffuser,
hanging from a float $\sim3$~m above the central PMT and illuminating all PMTs at the bottom of the tank.

The return light path, splitting off before the filter wheels, triggers a {\it Thorlabs} photo diode to generate a start
time record for each laser pulse and fans out into 150 separate optical fibers that go out to the connection boxes near
each tank pair. Inside each box, the return light path connects to a 30~m fiber, replicating the length of the fiber
connection between box and tank, and loops back through another 170~m fiber into the laser room.
Here, each channel can be selected individually via a {\it Dicon} switch to measure the laser pulse return time
with a {\it Hamamatsu} photomultiplier tube. The knowledge of these individual time constants of the detector array
provides the possibility to monitor even local variations, for example due to temperature dependent expansion or
contraction of fibers. A schematic of the laser system layout can be found in~\cite{bib:ICRC11cal}.

\section*{Calibration Method}

\subsection*{Data Taking}

For a calibration run, the system cycles through a wide range of optical filter wheel combinations that vary the laser
intensity over more than 4 orders of magnitude up to several thousand photo electrons (PE). 2000 laser pulses at
each intensity setting generate sufficient statistics for charge and slewing calibrations. The laser can be operated
with frequencies up to 500~Hz, but initial studies at the HAWC test WCD in Fort Collins, Colorado, US, showed that
operation at more than 200~Hz can lead to undesirable intensity variation. Using a frequency of 200~Hz and generating
2000 laser hits per filter wheel setting, a calibration cycle with 63 different intensities takes approximately
15~minutes.
This has to be repeated for 15 switch settings, each illuminating up to 10 tank pairs simultaneously. The duration of a
calibration run for the whole array will thus be less than 4~hours.
Furthermore, these runs do not increase the experiment's dead time significantly, since data taking is done with the
normal TDC DAQ which receives electronic trigger flags for the start time of each laser pulse.
Only a window of $~10 $~microseconds~($\mu s $) around the PMT response to each laser hit is tagged and excluded for air
shower reconstruction. Calibration runs are scheduled on a weekly basis and longterm studies of the calibration
stability will show if longer intervals in between are acceptable in future.

\subsection*{\label{timing} Timing Calibration}

The photo diode measurement of the laser pulse start time provides the trigger signal for the DAQ. A separate
{\it Berkeley Nucleonics} counter stores the time between this start signal and the second signal produced by the
light from the return loop. After averaging, half this time difference is a reliable measurement for the time between
the calibration trigger and the instance the laser photons reach the diffuser inside the tank. Subtracting this delay
and the time for traversing $\sim3$~m of water from the response time between trigger and TDC pulse measurement yields
the time offset due to electronics for each individual PMT. The correction of these time constants for each pulse time
that is part of an air shower event reduces timing mismatches and improves the angle fit of air shower fronts.

While not all fiber connections are installed, an alternative method to obtain time
offset corrections without laser signals is used. By reconstructing a large number of air showers, a systematic shift of
pulse times relative to the fitted shower front can be calculated for each individual PMT channel. The
details and results from this approach are discussed in the separate contribution~\cite{bib:ICRC13timecal}.

The exact relationship between the impact of a photon on the PMT cathode and the subsequent TDC record of an electronic
threshold crossing depends also on the pulse shape and its amplitude. A pulse with a higher charge, and therefore larger
ToT, has a shorter rise time and thus a reduced intrinsic delay, called slewing, for the TDC time record when
compared to a pulse of lower energy. Using the wide intensity range produced in a laser run, the relation between
either low threshold or high threshold ToT measurements and the slewing offset can be mapped as a
two-dimensional histogram. After subtracting the constant time offset, an empirical fit produces slewing
curves for both thresholds for each PMT. In the performance analysis in section~\ref{performance},
individual slewing results were not yet available for all active HAWC PMTs. Instead, average low and high threshold
slewing curves were used, derived from data collected at the HAWC test WCD. Details and examples of the slewing
procedure and results are presented in~\cite{bib:ICRC13timecal}.

\subsection*{\label{charge} Charge Calibration}

\begin{figure*}[!t]
  \centering
  \includegraphics[width=0.9\textwidth]{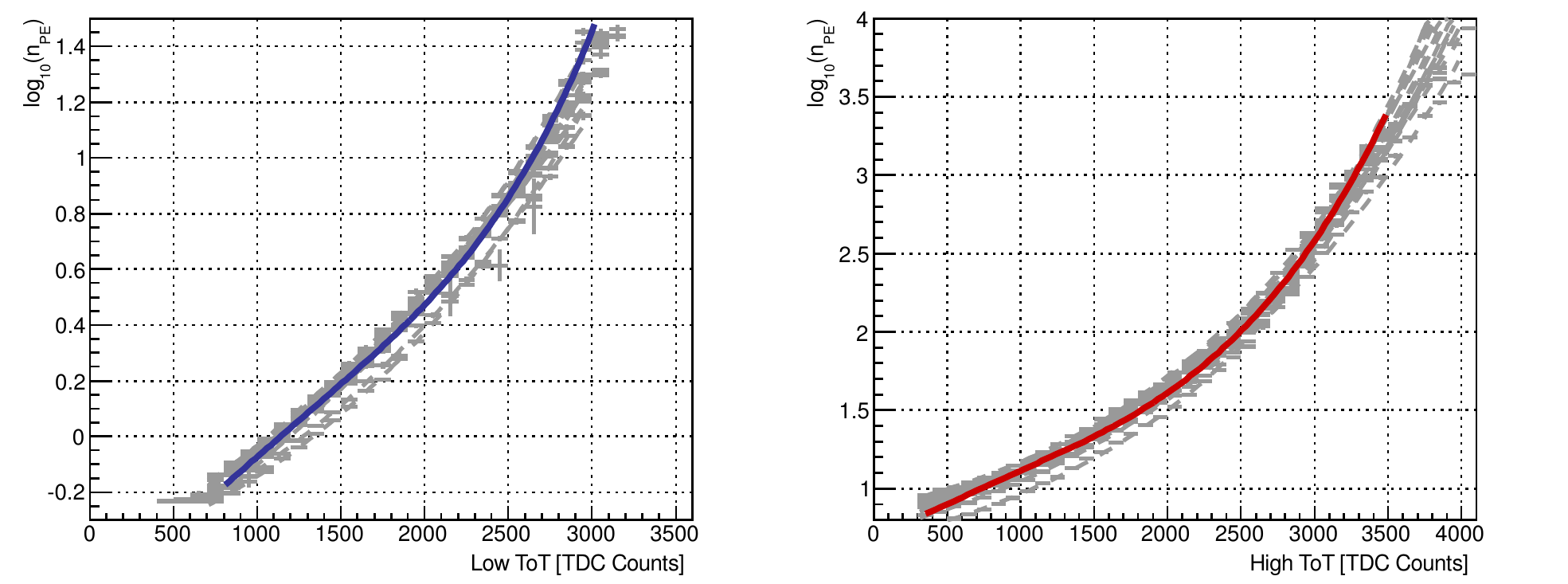}
  \caption{Examples of charge calibrations from one row of 4 WCDs in HAWC. The gray profile histograms and the gray
dotted lines are the individual low threshold (left) or high threshold (right) ToT to $\mbox{n}_{\mbox{\tiny PE}}$
conversion curves. The solid blue (left, low threshold) and solid red (right, high threshold) lines represent the
average curves derived at the test site and are the ones used for the performance study in section~\ref{performance}.}
  \label{fig::chargecurve}
 \end{figure*}

In HAWC, following the techniques used in the predecessor Milagro, the charge, and thus the weight, of individual
PMT pulses in an air shower event is measured as a ToT signal for two thresholds. Both values have to be calibrated
with the laser system.
Cycling through 63 different attenuation settings provides a wide laser intensity range that is monitored
with a {\it LaserProbe} radiometer. It receives light through a splitting cube in the light-to-tanks path and
is usually operated in a mode that averages over the 2000 laser pulses at a fixed filter wheel setting.
The occupancy $\eta=m/n$ for each laser intensity setting is obtained by counting the number of TDC responses $m$ in the
predetermined coincidence time window after the trigger signal, divided by the number of laser pulses $n$. The averaged
occupancy at a fixed intensity $\lambda$ in PE is equivalent to the Poisson probability of producing at least a
$1$~PE pulse,
$\eta = 1 - \exp(-\lambda)$ .
By substituting ${\lambda = a \cdot r_i/r_n}$, the measured occupancies are expressed as a function of the laser
radiometer measurements $r_i$, normalized by an arbitrary value $r_n$. A fit of the occupancies to low intensities
$\leq \sim2$~PE is performed to obtain the $a$ parameter, avoiding the regime where the uncertainty on
$\lambda$ diverges. The conversion factor $a$ can then be used to translate any relative radiometer measurement into a
mean PE intensity $\lambda$.
A Poisson distribution for each such $\lambda$, smeared with the energy resolution of the PMTs, produces
an expected distribution of numbers of PE ($\mbox{n}_{\mbox{\tiny PE}}$) that can be matched with quantiles of a
histogram of the TDC ToT measurements at that intensity. For either the low or the high threshold response, all these
$\mbox{n}_{\mbox{\tiny PE}}$-ToT pairs are merged together in profile histograms and fitted with an empirical function.
Using the calibration data from cycling through the full intensity spectrum of $<0.1$~PE to $>1000$~PE (exact values are
channel dependent) thus yields individual low and high threshold charge calibration curves for each PMT. 

The fitted ToT-PE conversions are applied analytically to every PMT pulse that is part of a triggered air shower event
to provide the calibrated charge as an accurate weight for the reconstruction algorithms. Whenever a pulse crosses the
high threshold, the corresponding ToT is the more reliable charge estimator and is used instead of the low threshold
ToT.
All PMTs were characterized before installation at the site and are grouped in batches with voltages chosen to
produce
approximately the same gain of $1.4\cdot 10^7$. This procedure also results in charge calibration curves that are
generally well aligned when comparing individual PMTs, see Fig.~\ref{fig::chargecurve} for some examples. This
similarity allows for use of an average charge calibration curve, derived from data collected at
the HAWC
WCD test site, that will be replaced with individual curves when the completion of the optical fiber network
allows for a calibration of all PMTs at the HAWC site. The average charge calibration was used for the performance analysis
shown in section~\ref{performance}.

\section*{\label{performance} Calibration Performance}

\begin{figure*}[!t]
  \centering
  \includegraphics[width=0.85\textwidth]{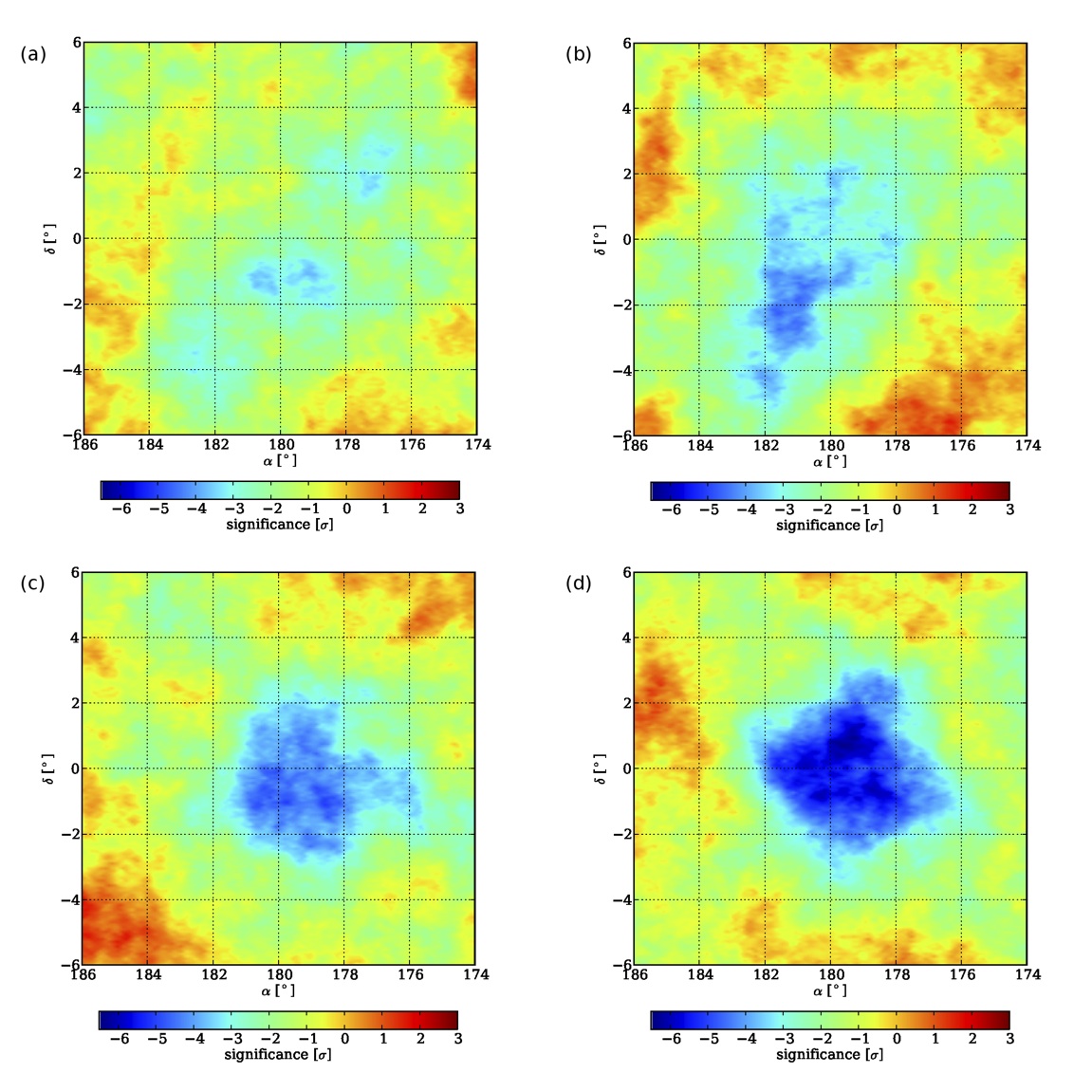}
  \caption{Significance maps of air shower event directions centered on the
  position of the moon, with $\alpha =$ event right ascension (RA) $-$ moon RA
  $+ 180^{\circ}$ and $\delta = $ event declination $-$ moon declination.  Map
  (a) is based on reconstructions with all pulse charges set to 1~PE and no
  timing calibration, map (b) has only charge calibrations included, map (c)
  has slewing and time residual calibrations applied but all charges are 1~PE,
  and map (d) was produced with full charge and timing calibrations included.}
  \label{fig:maps}
 \end{figure*}

The first data collected with 30 HAWC WCDs has limited sensitivity and statistics for gamma ray observations, but a
verification of the calibration performance can be achieved by mapping air shower event directions, dominated by charged
primary particles, around the position of the moon. The moon blocks these cosmic rays and
produces a deficit. A detailed analysis of this observation is presented in~\cite{bib:ICRC13moon}. Here, a qualitative
comparison of the position and shape of the moon shadow deficit for data from 30 HAWC WCDs with different
processing conditions highlights the improvement of event reconstruction through laser calibration.
A subset of data collected with the partial HAWC array between September an December 2012 with a total live time of 24.8
days was processed with four different calibration settings:
\begin{itemize}
 \item[(a)] All charges set to 1~PE and no timing calibration;
 \item[(b)] Charge calibration applied with average curve;
 \item[(c)] Timing calibration applied with average slewing curve and individual shower residual offsets, all charges
set to 1~PE;
 \item[(d)] Charge calibration as in (b) and timing calibration as in (c) applied.
\end{itemize}
For these data sets, all air shower events were reconstructed. Besides
requiring that the angle fit succeeded, the only cut applied to the results was
removing events with less than 32 PMT channels participating, to exclude low
energy showers that reduce the deficit significance.  To correct for a known
inaccuracy in the survey of PMT positions, an overall shift was applied to all
direction results, chosen in such a way that the maximum of the angular shower
distribution realigns with the local zenith.  A check was performed to confirm
that this correction slightly increases the deficit significance but does not
strongly affect the shape of the moon shadow in any of the four cases.  The
results for the four cases are shown in Fig.~\ref{fig:maps} as maps of binned
statistical significances with a radial smoothing of $3^{\circ}$ applied.  In
the first map (a), no unambiguous moon shadow is visible and no deficit with a
significance of at least $-5.0 \sigma$ can be found.  After applying only the
charge calibration (b), a deficit (peak significance $-5.0 \sigma$) is visible
but is both of irregular shape and offset from the moon position by several
degrees. Fig.~\ref{fig:maps}~(c) shows that using only the timing calibration
and no charge calibration deepens the deficit only slightly ($-5.1 \sigma$) but
produces a clearer image of the moon shadow closer to the actual location. This
was expected due to the importance of individual pulse times for fitting the
shower front. The final comparison with the map (d) in which both charge and
timing calibrations are applied reveals a much more pronounced deficit ($-6.4
\sigma$) and a more symmetric and well centered moon shadow. Counting events
over the full sky coverage, the last map contains $\sim 1.1 \cdot 10^9$ events
and thus $\sim 3 \cdot 10^8$ more than those maps from cases (a), (b) and (c),
due to the fact that more shower angle fits are successful when full
calibrations are used.

\section*{Conclusions}

The main components of the HAWC laser calibration system are installed and
operational and calibration runs can be performed without significantly
reducing the array's dead time. The experience gained from a WCD test setup and
systematic time residuals derived from shower fits made it possible to
calibrate early HAWC data even before calibration results for all individual
PMT channels are available. Both charge and timing calibrations significantly
improve air shower reconstruction as is shown here based on the mapping of the
moon shadow deficit for a subset of early HAWC data.  The optical fiber network
will continue to grow with the HAWC array and provides the means for regular
calibrations of all PMTs to guarantee a strong and stable performance in
gamma-ray observations.

\vspace*{0.5cm}

\section*{Acknowledgments}

We acknowledge the support from: US National Science Foundation (NSF); US
Department of Energy Office of High-Energy Physics; The Laboratory Directed
Research and Development (LDRD) program of Los Alamos National Laboratory;
Consejo Nacional de Ciencia y Tecnolog\'{\i}a (CONACyT), M\'exico; Red de
F\'{\i}sica de Altas Energ\'{\i}as, M\'exico; DGAPA-UNAM, M\'exico; and the
University of Wisconsin Alumni Research Foundation.

% 
% \vspace*{0.5cm}
% \footnotesize{{\bf Acknowledgment:}{The ICRC 2013 is funded by FAPERJ, CNPq, FAPESP, CAPES and IUPAP.}}
% 

\clearpage

%\end{document}


\begin{thebibliography}{}

\bibitem{bib:Miguel} M. Mostafa, Session HL of ICRC 2013
\bibitem{bib:Hao} H. Zhao and H. Ayala, Session GA-IN Poster 651 of ICRC 2013
\bibitem{bib:Ibrahim} I. Torres, Session GA-IN Poster 824 of ICRC 2013

\end{thebibliography}

\begin{thebibliography}{}

\bibitem{bib:hawc} Abeysekara, U.A., et al., ÒOn the sensitivity of the HAWC observatory to gamma-ray burstsÓ, Astroparticle Physics
35, 641Ð650 (2012).

\bibitem{bib:milagro} Atkins R., et al., ÒObservation of TeV gamma rays from the Crab Nebula with MILAGRO using a new background
rejection techniqueÓ, Astrophysical Journal 595, 803-811 (2003).


\end{thebibliography}

\begin{thebibliography}{}


\bibitem{bib:hawcSensi} Abeysekara, U.A., et al., On the sensitivity of the HAWC
observatory to gamma-ray bursts, Astroparticle Physics 35, 641:650 (2012).

\bibitem{bib:pretz} J. Pretz and J. Goodman these proceedings.

\bibitem{bib:ARGO} Aielli, G., Assiro, R., Bacci, C., et al. 2006, Nucl.
Instrum. Methods Phys. Res. A, 562, 92.

\bibitem{bib:HISCORE} M.Tluczykont et al , ArXiv: 0909.0445

\end{thebibliography}

\begin{thebibliography}{}

\bibitem{icrc2007} Carrami\~nana, A., Gonz\'alez, M.M., Salazar, H., et al. Proc. of the 30th ICRC 3, 1567 (2008).

\bibitem{hawc-moon} Fiorino, D. for the HAWC Collaboration, ICRC 2013, contrib. 784.

\bibitem{anisotropy} BenZvi, S.. for the HAWC Collaboration, ICRC 2013, contrib. 710.

%\bibitem{bib:lattes} C.M.G. Lattes, G.P.S. Occhialini and  C.F. Powell, Nature 160 (1947) 486-492 doi:10.1038/160486a0.

%\bibitem{bib:schoenberg} M. Sch\"onberg and S. Chandrasekhar, The Astrophysical Journal 96 (1942) 161-172 doi:10.1086/144444.

\end{thebibliography}

\begin{thebibliography}{}

\bibitem{bib:Mostafa} Mostaf\'{a}, M. 2013, 33rd International Cosmic Ray
Conference, The HAWC Observatory.

\bibitem{bib:Calibration} Huentemeyer, P. et al., ICRC2011, Beijing, China. (2011)

\bibitem{bib:HaoSlewing} Zhou, H. Slewing calibration using the CSU WCD prototype. HAWC Collaboration Technical Note. (2012)

\bibitem{bib:abdo} Abdo, A.A. Discovery of localized TeV gamma-ray sources and diffuse TeV gamma-ray emission from the galactic plane with Milagro using a new rejection technique. (2007)

\bibitem{bib:lansdell} Lansdell C. Timing pedestal shifting and the Crab. Milagro Collaboration Memorandum. (2005) 

\bibitem{bib:lauer} Lauer, R. 2013, 33rd International Cosmic Ray Conference, Calibration and Reconstruction Performance of the HAWC Observatory.

\end{thebibliography}

\begin{thebibliography}{}

\bibitem{bib:Milagro} R. Atkins et al., Astrophys. J., {\bf 608} 680 (2004).
%\bibitem{bib:laser} www.teemphotonics.com .
\bibitem{bib:HAWCgrb} A. Abeysekara et al. (HAWC Collaboration), Astropart. Phys., {\bf 35} 641 (2012).
\bibitem{bib:ICRC13general} M. Mostafa et al. (HAWC Collaboration), these proc.
\bibitem{bib:ICRC13sensi} J. Pretz et al. (HAWC Collaboration), these proc.
\bibitem{bib:ICRC11cal} P. Huentmeyer et al. (HAWC Collaboration), Proc. of ICRC 2011 doi:10.7529/ICRC2011/V09/0767.
\bibitem{bib:ICRC13timecal} H. Ayala Solares et al. (HAWC Collaboration), these proc.
\bibitem{bib:ICRC13moon} D. Fiorino et al. (HAWC Collaboration), these proc.

\end{thebibliography}
\end{document}